\documentclass[12pt,a4paper]{article}

\usepackage[margin=1in]{geometry}

\usepackage{graphicx}  % to include figures (can also use other packages)

\usepackage{xspace}
\usepackage{color}
\usepackage{colortbl}
\usepackage{subfigure}
\usepackage{amsmath}

\usepackage{ifthen} % for conditional statements

\newboolean{pdflatex}
\setboolean{pdflatex}{false} % use this if using eps figures
\newboolean{uprightparticles}
\setboolean{uprightparticles}{false} %Set to true to get roman particle symbols
\usepackage{amssymb}
\usepackage{amsfonts}
\usepackage{upgreek}
\usepackage{textcomp}
\usepackage{multirow}
\usepackage{booktabs}
\usepackage[nottoc,numbib]{tocbibind}
\usepackage[toc,page]{appendix}
\usepackage{hyperref}
\hypersetup{%
    colorlinks=true, linktocpage=true, pdfstartpage=3, pdfstartview=FitV,%
   breaklinks=true, pdfpagemode=UseNone, pageanchor=true, pdfpagemode=UseOutlines,%
    plainpages=false, bookmarksnumbered, bookmarksopen=true, bookmarksopenlevel=1,%
    hypertexnames=true, pdfhighlight=/O,%
    urlcolor=blue, linkcolor=blue,% 
}
\def\lhcb {LHCb\xspace}
\def\ux85 {UX85\xspace}

\def\lhc {LHC\xspace}

\def\cms {CMS\xspace}

\ifthenelse{\boolean{uprightparticles}}%
{

 \def\Ppi         {\ensuremath{\uppi}\xspace}

 \def\PDelta      {\ensuremath{\Delta}\xspace}                 
 \def\PXi      {\ensuremath{\Xi}\xspace}                 
 \def\PLambda      {\ensuremath{\Lambda}\xspace}                 
 \def\PSigma      {\ensuremath{\Sigma}\xspace}                 
 \def\POmega      {\ensuremath{\Omega}\xspace}                 
 \def\PUpsilon      {\ensuremath{\Upsilon}\xspace}                 
 
 \def\PB      {\ensuremath{\mathrm{B}}\xspace}                 
                  
 \def\PD      {\ensuremath{\mathrm{D}}\xspace}

 \def\PK      {\ensuremath{\mathrm{K}}\xspace}

 \def\PV      {\ensuremath{\mathrm{V}}\xspace}

 \def\Pc      {\ensuremath{\mathrm{c}}\xspace}

 \def\Pi      {\ensuremath{\mathrm{i}}\xspace}

 \def\Pp      {\ensuremath{\mathrm{p}}\xspace}

}
{

 \def\Ppi         {\ensuremath{\pi}\xspace}

 \mathchardef\PDelta="7101
 \mathchardef\PXi="7104
 \mathchardef\PLambda="7103
 \mathchardef\PSigma="7106
 \mathchardef\POmega="710A
 \mathchardef\PUpsilon="7107
                  
 \def\PB      {\ensuremath{B}\xspace}                 
                  
 \def\PD      {\ensuremath{D}\xspace}

 \def\PK      {\ensuremath{K}\xspace}

 \def\PV      {\ensuremath{V}\xspace}

 \def\Pc      {\ensuremath{c}\xspace}

 \def\Pi      {\ensuremath{i}\xspace}

 \def\Pp      {\ensuremath{p}\xspace}

}

\def\s     {\ensuremath{\Ps}\xspace}

\def\c     {\ensuremath{\Pc}\xspace}

\def\pion  {\ensuremath{\Ppi}\xspace}

\def\pip   {\ensuremath{\pion^+}\xspace}
\def\pim   {\ensuremath{\pion^-}\xspace}
\def\pipi  {\ensuremath{\pion^+\pion^-}\xspace}

\def\kaon  {\ensuremath{\PK}\xspace}
\def\Kbar  {\kern 0.2em\overline{\kern -0.2em \PK}{}\xspace}

\def\Kz    {\ensuremath{\kaon^0}\xspace}
\def\Kzb   {\ensuremath{\Kbar^0}\xspace}
\def\KzKzb {\ensuremath{\Kz \kern -0.16em \Kzb}\xspace}
\def\Kp    {\ensuremath{\kaon^+}\xspace}
\def\Km    {\ensuremath{\kaon^-}\xspace}

\def\KpKm  {\ensuremath{\Kp \kern -0.16em \Km}\xspace}
\def\KS    {\ensuremath{\kaon^0_{\rm\scriptscriptstyle S}}\xspace}

\def\Dbar    {\kern 0.2em\overline{\kern -0.2em \PD}{}\xspace}
\def\D       {\ensuremath{\PD}\xspace}

\def\Dz      {\ensuremath{\D^0}\xspace}
\def\Dzb     {\ensuremath{\Dbar^0}\xspace}
\def\DzDzb   {\ensuremath{\Dz {\kern -0.16em \Dzb}}\xspace}
\def\Dp      {\ensuremath{\D^+}\xspace}
\def\Dm      {\ensuremath{\D^-}\xspace}

\def\DpDm    {\ensuremath{\Dp {\kern -0.16em \Dm}}\xspace}

\def\Bbar    {\kern 0.18em\overline{\kern -0.18em \PB}{}\xspace}

\def\Y#1S{\ensuremath{\PUpsilon{(#1S)}}\xspace}% no space before {...}!

\def\proton      {\ensuremath{\Pp}\xspace}
\def\antiproton  {\ensuremath{\overline \proton}\xspace}

\def\L {\ensuremath{\PLambda}\xspace}
\def\Lbar{\ensuremath{\overline \L}\xspace}

\newcommand{\decay}[2]{\ensuremath{#1\!\to #2}\xspace}         % {\Pa}{\Pb \Pc}

\def\to                 {\ensuremath{\rightarrow}\xspace}

\def\AT#1     {\ensuremath{A_T^{#1}}\xspace}           % 2

\def\C#1      {\ensuremath{\mathcal{C}_{#1}}\xspace}                       % 9
\def\Cp#1     {\ensuremath{\mathcal{C}_{#1}^{'}}\xspace}                    % 7
\def\Ceff#1   {\ensuremath{\mathcal{C}_{#1}^{\mathrm{(eff)}}}\xspace}        % 9  
\def\Cpeff#1  {\ensuremath{\mathcal{C}_{#1}^{'\mathrm{(eff)}}}\xspace}       % 7
\def\Ope#1    {\ensuremath{\mathcal{O}_{#1}}\xspace}                       % 2
\def\Opep#1   {\ensuremath{\mathcal{O}_{#1}^{'}}\xspace}                    % 7

\newcommand{\tev}{\ensuremath{\mathrm{\,Te\kern -0.1em V}}\xspace}
\newcommand{\gev}{\ensuremath{\mathrm{\,Ge\kern -0.1em V}}\xspace}
\newcommand{\mev}{\ensuremath{\mathrm{\,Me\kern -0.1em V}}\xspace}
\newcommand{\kev}{\ensuremath{\mathrm{\,ke\kern -0.1em V}}\xspace}
\newcommand{\ev}{\ensuremath{\mathrm{\,e\kern -0.1em V}}\xspace}
\newcommand{\gevc}{\ensuremath{{\mathrm{\,Ge\kern -0.1em V\!/}c}}\xspace}
\newcommand{\mevc}{\ensuremath{{\mathrm{\,Me\kern -0.1em V\!/}c}}\xspace}
\newcommand{\gevcc}{\ensuremath{{\mathrm{\,Ge\kern -0.1em V\!/}c^2}}\xspace}
\newcommand{\gevgevcccc}{\ensuremath{{\mathrm{\,Ge\kern -0.1em V^2\!/}c^4}}\xspace}
\newcommand{\mevcc}{\ensuremath{{\mathrm{\,Me\kern -0.1em V\!/}c^2}}\xspace}

\def\m    {\ensuremath{\rm \,m}\xspace}
\def\cm   {\ensuremath{\rm \,cm}\xspace}

\def\mm   {\ensuremath{\rm \,mm}\xspace}

\newcommand{\chisq}{\ensuremath{\chi^2}\xspace}

\def\gsim{{~\raise.15em\hbox{$>$}\kern-.85em
          \lower.35em\hbox{$\sim$}~}\xspace}
\def\lsim{{~\raise.15em\hbox{$<$}\kern-.85em
          \lower.35em\hbox{$\sim$}~}\xspace}

\def\pt         {\mbox{$p_T$}\xspace}

\def\mrad{\ensuremath{\rm \,mrad}\xspace}
\def\urad{\ensuremath{\textrm{\,\textmu rad}}\xspace}

\def\evtgen     {\mbox{\textsc{EvtGen}}\xspace}
\def\pythia     {\mbox{\textsc{Pythia}}\xspace}
\def\photos     {\mbox{\textsc{Photos}}\xspace}

\def\geant      {\mbox{\textsc{Geant\,4}}\xspace}

\def\tell1  {TELL1\xspace}
\def\ukl1   {UKL1\xspace}

\newcommand{\eg}{\mbox{e.g.}\xspace}
\newcommand{\ie}{\mbox{i.e.}\xspace}
\newcommand{\etal}{\mbox{et al.\/}\xspace}

\def\alice {ALICE\xspace}
\def\star {STAR\xspace}
\def\rhic {RHIC\xspace}

\def\VZ {\ensuremath{\PV^0}\!\xspace}

\def\Lbar{\ensuremath{\kern 0.2em\overline{\kern -0.2em\L\kern 0.05em}\kern-0.05em{}}\xspace}

\def\LToPpi {\ensuremath{\decay{\L}{\proton\pim}}\xspace}
\def\LbarToPpi {\ensuremath{\decay{\Lbar}{\antiproton\pip}}\xspace}
\def\KSTopipi {\ensuremath{\decay{\KS}{\pip\pim}}\xspace}

\def\pt {\ensuremath{p_\mathrm{T}}\xspace} % overwrites standard
\def\y {\ensuremath{y}\xspace}
\def\dy {\ensuremath{\Delta y}\xspace}
\def\FIP {\ensuremath{\mathcal{F}_{\mathrm{IP}}}\xspace}

\def\s {\ensuremath{\sqrt{s}}\xspace}
\def\pp {\ensuremath{\Pp\Pp}\xspace}
\def\LHCLow {\ensuremath{\s = 0.9\tev}\xspace}
\def\LHCHigh {\ensuremath{\s = 7\tev}\xspace}

\def\LintLow {\ensuremath{0.3\,\mathrm{nb}^{-1}}\xspace}
\def\LintHigh {\ensuremath{1.8\,\mathrm{nb}^{-1}}\xspace}

\def\LHCbUp {up\xspace}
\def\LHCbDown {down\xspace}

\begin{document}

%%%%%%%%%%%%%%%%%%%%%%%%%
%%%%% Title     %%%%%%%%%
%%%%%%%%%%%%%%%%%%%%%%%%%

%% -----------------------------------------------------------------
%% Measurement of V0 production ratios in pp collisions at _/s = 0.9 and 7 TeV
%% CERN-PH-EP-2011-082
%% -----------------------------------------------------------------
%% LHCb Collaboration 
%% T. Blake, C. Blanks, W. Bonivento, F. Dettori, R. Muresan
%% c.blanks07@imperial.ac.uk
%% -----------------------------------------------------------------

%%%%%%%%%%%%%%%%%%%%%%%%%
%%%%%  TITLE PAGE  %%%%%%
%%%%%%%%%%%%%%%%%%%%%%%%%
\begin{titlepage}
\pagenumbering{roman}

% Header ---------------------------------------------------
\vspace*{-1.5cm}
\centerline{\large EUROPEAN ORGANIZATION FOR NUCLEAR RESEARCH (CERN)}
\vspace*{1.5cm}
\hspace*{-0.5cm}
\begin{tabular*}{\linewidth}{lc@{\extracolsep{\fill}}r}
\ifthenelse{\boolean{pdflatex}}% Logo format choice
{\vspace*{-2.7cm}\mbox{\!\!\!\includegraphics[width=.14\textwidth]{figs/lhcb-logo.pdf}} & &}%
{\vspace*{-1.2cm}\mbox{\!\!\!\includegraphics[width=.12\textwidth]{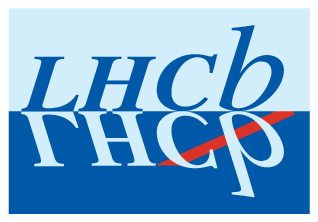}} & &}%
\\
 & & LHCb-PAPER-2011-005 \\  % ID 
 & & CERN-PH-EP-2011-082 \\  % ID 
& & \today \\ 
 & & \\
\end{tabular*}

\vspace*{4.0cm}

% Title --------------------------------------------------
{\bf\boldmath\huge
\begin{center}
  Measurement of \VZ production ratios in \pp collisions at \mbox{$\s =$ 0.9 and 7\tev}
\end{center}
}

\vspace*{2.0cm}

% Authors -------------------------------------------------
\begin{center}
The LHCb Collaboration\,\footnote{Authors are listed on the following pages.}
\end{center}

\vspace{\fill}

% Abstract -----------------------------------------------
\begin{abstract}
\noindent The \Lbar/\L and \Lbar/\KS production ratios are measured by the \lhcb detector from \LintLow of \pp collisions delivered by the \lhc at \LHCLow and \LintHigh at \LHCHigh.  Both ratios are presented as a function of transverse momentum, \pt, and rapidity, \y, in the ranges \mbox{$0.15<\pt<2.50\gevc$} and \mbox{$2.0<\y<4.5$}.  Results at the two energies are in good agreement as a function of rapidity loss, $\dy = y_{\textrm{beam}} - y$, and are consistent with previous measurements. The ratio \Lbar/\L, measuring the transport of baryon number from the collision into the detector, is smaller in data than predicted in simulation, particularly at high rapidity.  The ratio \Lbar/\KS, measuring the baryon-to-meson suppression in strange quark hadronisation, is significantly larger than expected.
\end{abstract}

\vspace*{2.0cm}
\vspace{\fill}

\end{titlepage}

%%%%%%%%%%%%%%%%%%%%%%%%%%%%%%%%
%%%%%  EOD OF TITLE PAGE  %%%%%%
%%%%%%%%%%%%%%%%%%%%%%%%%%%%%%%%

%  empty page follows the title page ----
\newpage
\setcounter{page}{2}
\mbox{~}
\newpage

% Author List ----------------------------
% \documentclass[a4paper]{article}
% \setlength{\oddsidemargin}{0cm}
% \setlength{\evensidemargin}{0cm}
% \setlength{\textwidth}{16.5cm}
% \setlength{\parindent}{0cm}
% \begin{document}
% \begin{flushleft}
% {\Large LHCb Collaboration ----- official authorship list}\\[4ex]
% valid for date: 1. Jun. 2011\\
% used for paper: Authorlist for June 2011\\[4ex]
% collaborators included, who did not leave before 1. Jun. 2010\\
%                            and who joined before 1. Dec. 2010\\[2ex]
% {\small today is 2. Jun. 2011}\\[4ex]
%-- 
%-- LHCb Authorlist, Status of 1. Jun. 2011
%-- 
\centerline{\large\bf The LHCb Collaboration}
\begin{flushleft}
\small
R.~Aaij$^{23}$, 
B.~Adeva$^{36}$, 
M.~Adinolfi$^{42}$, 
C.~Adrover$^{6}$, 
A.~Affolder$^{48}$, 
Z.~Ajaltouni$^{5}$, 
J.~Albrecht$^{37}$, 
F.~Alessio$^{6,37}$, 
M.~Alexander$^{47}$, 
G.~Alkhazov$^{29}$
% ...

{\it
% ...
$ ^{2}$Universidade Federal do Rio de Janeiro (UFRJ), Rio de Janeiro, Brazil\\
% ...
$ ^{4}$LAPP, Universit\'{e} de Savoie, CNRS/IN2P3, Annecy-Le-Vieux, France\\
$ ^{5}$Clermont Universit\'{e}, Universit\'{e} Blaise Pascal, CNRS/IN2P3, LPC, Clermont-Ferrand, France\\
$ ^{6}$CPPM, Aix-Marseille Universit\'{e}, CNRS/IN2P3, Marseille, France\\
% ...
$ ^{15}$Sezione INFN di Cagliari, Cagliari, Italy\\
% ...
$ ^{23}$Nikhef National Institute for Subatomic Physics, Amsterdam, Netherlands\\
% ...
$ ^{29}$Petersburg Nuclear Physics Institute (PNPI), Gatchina, Russia\\
% ...
$ ^{36}$Universidad de Santiago de Compostela, Santiago de Compostela, Spain\\
$ ^{37}$European Organization for Nuclear Research (CERN), Geneva, Switzerland\\
% ...
$ ^{42}$H.H. Wills Physics Laboratory, University of Bristol, Bristol, United Kingdom\\
% ...
$ ^{47}$School of Physics and Astronomy, University of Glasgow, Glasgow, United Kingdom\\
$ ^{48}$Oliver Lodge Laboratory, University of Liverpool, Liverpool, United Kingdom\\
% ...
$ ^{50}$School of Physics and Astronomy, University of Manchester, Manchester, United Kingdom\\
% ...
}
\bigskip
% ---- LHCb Authorlist, Status 1. Jun. 2011
% ---- Number of Authors = 557
% ---- 
\end{flushleft}
% \end{document}

\cleardoublepage

%%%%%%%%%%%%%%%%%%%%%%%%%
%%%%% Main text %%%%%%%%%
%%%%%%%%%%%%%%%%%%%%%%%%%

\pagestyle{plain} % restore page numbers for the main text
\setcounter{page}{1}
\pagenumbering{arabic}

%% -----------------------------------------------------------------
%% Measurement of V0 production ratios in pp collisions at _/s = 0.9 and 7 TeV
%% CERN-PH-EP-2011-082
%% -----------------------------------------------------------------
%% LHCb Collaboration 
%% T. Blake, C. Blanks, W. Bonivento, F. Dettori, R. Muresan
%% c.blanks07@imperial.ac.uk
%% -----------------------------------------------------------------

\section{Introduction}
\label{sec:introduction}

While the underlying interactions of hadronic collisions and hadronisation are understood within the Standard Model, exact computation of the processes governed by QCD are difficult due to the highly non-linear nature of the strong force.  In the absence of full calculations, generators based on phenomenological models have been devised and optimised, or ``tuned'', to accurately reproduce experimental observations.  These generators predict how Standard Model physics will behave at the \lhc and constitute the reference for discoveries of New Physics effects.  

Strange quark production is a powerful probe for hadronisation processes at \pp colliders since protons have no net strangeness.  Recent experimental results in the field have been published by \star~\cite{STAR} from \rhic \pp collisions at $\s = 0.2\tev$ and by \alice~\cite{ALICE}, \cms~\cite{CMS} and \lhcb~\cite{KsCrossSection} from \lhc \pp collisions at $\s = 0.9$ and 7\tev.  \lhcb can make an important contribution thanks to a full instrumentation of the detector in the forward region that is unique among the \lhc experiments.  Studies of data recorded at different energies with the same apparatus help to control the experimental systematic uncertainties.  

In this paper we report on measurements of the efficiency corrected production ratios of the strange particles \Lbar, \L and \KS as observables related to the fundamental processes behind parton fragmentation and hadronisation.  The ratios

\begin{align}
\frac{\Lbar}{\L} &=\frac{\sigma(\decay{\pp}{\Lbar X})}{\sigma(\decay{\pp}{\L X})} 
\intertext{and}
\frac{\Lbar}{\KS} &=\frac{\sigma(\decay{\pp}{\Lbar X})}{\sigma(\decay{\pp}{\KS X})}
\end{align}
\noindent have predicted dependences on rapidity, \y, and transverse momentum, \pt, which can vary strongly between different tunes of the generators.  

Measurements of the ratio \Lbar/\L allow the study of the transport of baryon number from \pp collisions to final state hadrons and the ratio \Lbar/\KS is a measure of baryon-to-meson suppression in strange quark hadronisation.

%% -----------------------------------------------------------------
%% Measurement of V0 production ratios in pp collisions at _/s = 0.9 and 7 TeV
%% CERN-PH-EP-2011-082
%% -----------------------------------------------------------------
%% LHCb Collaboration 
%% T. Blake, C. Blanks, W. Bonivento, F. Dettori, R. Muresan
%% c.blanks07@imperial.ac.uk
%% -----------------------------------------------------------------

\section{The LHCb detector and data samples}
\label{sec:detector}

The Large Hadron Collider beauty experiment (LHCb) at CERN is a single-arm spectrometer covering the forward rapidity region.  The analysis presented in this paper relies exclusively on the tracking detectors.  The high precision tracking system begins with a silicon strip Vertex Locator (VELO), designed to identify displaced secondary vertices up to about 65\cm downstream of the nominal interaction point.  A large area silicon tracker follows upstream of a dipole magnet and tracker stations, built with a mixture of straw tube and silicon strip detectors, are located downstream.  The \lhcb coordinate system is defined to be right-handed with its origin at the nominal interaction point, the $z$ axis aligned along the beam line towards the magnet and the $y$ axis pointing upwards.  The bending plane is horizontal and the magnet has a reversible field, with the positive $B_y$ polarity called ``up'' and the negative ``down''.  Tracks reconstructed through the full spectrometer experience an integrated magnetic field of around 4\,Tm.  The detector is described in full elsewhere \cite{DetectorPaper}.

A loose minimum bias trigger is used for this analysis, requiring at least one track segment in the downstream tracking stations.  This trigger is more than $99$\,\% efficient for offline selected events that contain at least two tracks reconstructed through the full system.
 
Complementary data sets were recorded at two collision energies of $\s = 0.9$ and 7\tev, with both polarities of the dipole magnet.  An integrated luminosity of \LintLow (corresponding to 12.5 million triggers) was taken at the lower energy, of which 48\,\% had the \LHCbUp magnetic field configuration.  At the higher energy, 67\,\% of a total \LintHigh (110.3 million triggers) was taken with field \LHCbUp.

At injection energy (\LHCLow), the proton beams are significantly broadened spatially compared to the accelerated beams at \LHCHigh.  To protect the detector, the two halves of the VELO are retracted along the $x$ axis from their nominal position of inner radius of 8\mm to the beam, out to 18\mm, which results in a reduction of the detector acceptance at small angles to the beam axis by approximately 0.5 units of rapidity.

The beams collide with a crossing angle in the horizontal plane tuned to compensate for \lhcb's magnetic field.  The angle required varies as a function of beam configuration and for the data taking period covered by this study was set to 2.1\mrad at \LHCLow and 270\urad at 7\tev.  Throughout this analysis \VZ momenta and any derived quantity such as rapidity are computed in the centre-of-mass frame of the colliding protons.

Samples of Monte Carlo (MC) simulated events have been produced in close approximation to the data-taking conditions described above for estimation of efficiencies and systematic uncertainties.  A total of 73 million simulated minimum bias events were used for this analysis per magnet polarity at \LHCLow and 60\,(69) million events at 7\tev for field \LHCbUp\,(\LHCbDown).  \lhcb MC simulations are described in Ref.~\cite{Gauss}, with \pp collisions generated by \pythia\!6\,\cite{PYTHIA6}.  Emerging particles decay via \evtgen\!\cite{EvtGen}, with final state radiation handled by \photos\!\cite{PHOTOS}.  The resulting particles are transported through \lhcb by \geant\!\cite{Geant}, which models hits on the sensitive elements of the detector as well as interactions between the particles and the detector material.  Secondary particles produced in these material interactions decay via \geant.

Additional samples of five million minimum bias events were generated for studies of systematic uncertainties using \pythia\!6 variants Perugia\,0 (tuned on experimental results from SPS, LEP and Tevatron) and Perugia\,NOCR (an extreme model of baryon transport) \cite{Peter}.  Similarly sized samples of \pythia\!8\,\cite{PYTHIA8} minimum bias diffractive events were also generated, including both hard and soft diffraction\,\footnote{Single- and double-diffractive process types are considered: 92--94 in \pythia\!6.421, with soft diffraction, and 103--105 in \pythia\!8.130, with soft and hard diffraction.}\,\cite{Navin}.

%% -----------------------------------------------------------------
%% Measurement of V0 production ratios in pp collisions at _/s = 0.9 and 7 TeV
%% CERN-PH-EP-2011-082
%% -----------------------------------------------------------------
%% LHCb Collaboration 
%% T. Blake, C. Blanks, W. Bonivento, F. Dettori, R. Muresan
%% c.blanks07@imperial.ac.uk
%% -----------------------------------------------------------------

\section{Analysis procedure}
\label{sec:analysis}

\VZ hadrons are named after the ``V''-shaped track signature of their dominant decays: \LToPpi, \LbarToPpi and \KSTopipi, which are reconstructed for this analysis.  Only tracks with quality $\chisq / \textrm{ndf} <9$ are considered, with the \VZ required to decay within the VELO and the daughter tracks to be reconstructed through the full spectrometer.  Any oppositely-charged pair is kept as a potential \VZ candidate if it forms a vertex with $\chisq < 9$ (with one degree of freedom for a \VZ vertex).  \Lbar, \L and \KS candidates are required to have invariant masses within $\pm50$\mevcc of the PDG values\,\cite{PDG}.  This mass window is large compared to the measured mass resolutions of about 2\mevcc for \L\!(\Lbar) and 5\mevcc for \KS.

\begin{figure}
  \centering
  \subfigure[]{
    \includegraphics[width=0.47\textwidth]{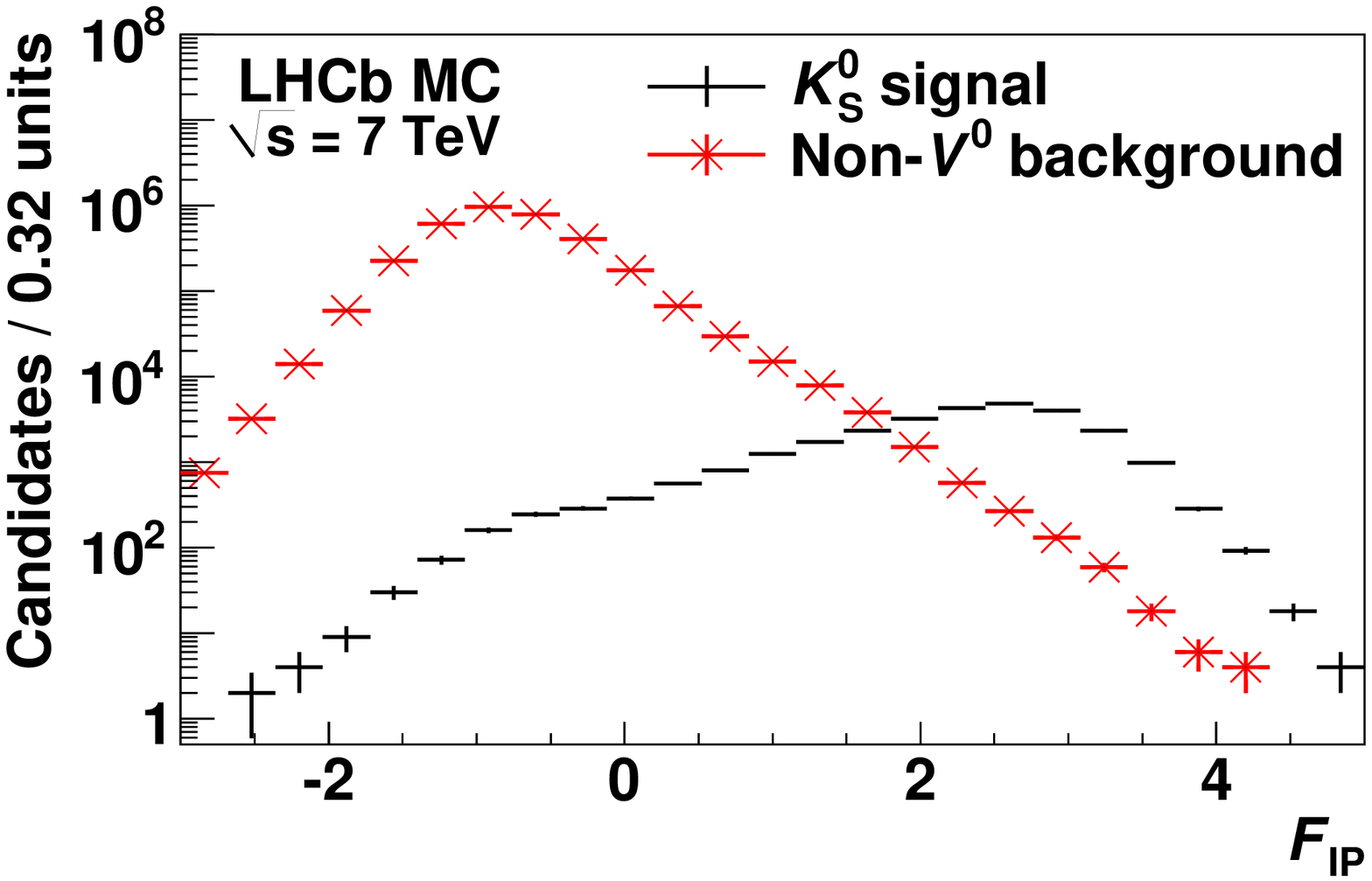}
      \label{fig:fisher-k}
  }
  \subfigure[]{
    \includegraphics[width=0.47\textwidth]{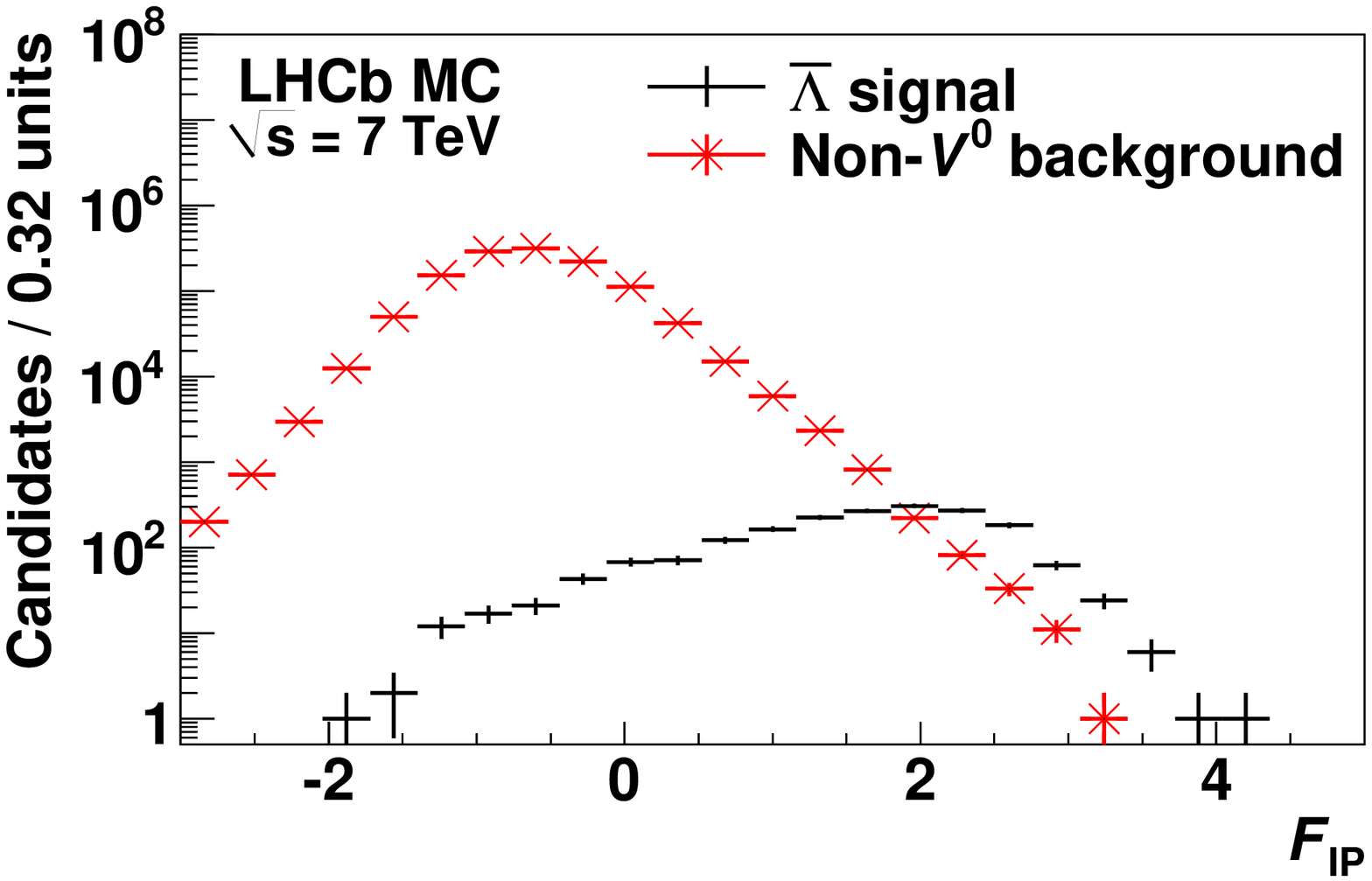}
      \label{fig:fisher-a}
  }
  \caption{\small The Fisher discriminant \FIP in 0.5\,million Monte Carlo simulated minimum bias events at \LHCHigh for \subref{fig:fisher-k} \KS and \subref{fig:fisher-a} \Lbar.}
  \label{fig:fisher}
\end{figure}

Combinatorial background is reduced with a Fisher discriminant based on the impact parameters (IP) of the daughter tracks ($d^{\pm}$) and of the reconstructed \VZ mother, where the impact parameter is defined as the minimum distance of closest approach to the nearest reconstructed primary interaction vertex measured in mm.  The Fisher discriminant:
\begin{equation}
\FIP = a \log_{10}(d^{+}_\mathrm{IP}/1\,\mathrm{mm}) + b \log_{10}(d^{-}_\mathrm{IP}/1\,\mathrm{mm}) + c \log_{10}(\VZ_\mathrm{IP}/1\,\mathrm{mm})
\end{equation}
\noindent is optimised for signal significance ($S/\sqrt{S+B}$) on simulated events after the above quality criteria.  The cut value, $\FIP>1$, and coefficients, $a=b=-c=1$, were found to be suitable for \Lbar, \L and \KS at both collision energies (Fig.~\ref{fig:fisher}).

The \L\!(\Lbar) signal significance is improved by a $\pm4.5$\mevcc veto around the PDG \KS mass after re-calculation of each candidate's invariant mass with an alternative \pipi daughter hypothesis.  A similar veto to remove \L\!(\Lbar) with a $\proton\pim$\,($\antiproton\pip$) hypothesis from the \KS sample is not found to improve significance so is not applied.

\begin{figure}[t!]
\centering
\subfigure[]{
  \includegraphics[width=0.47\textwidth]{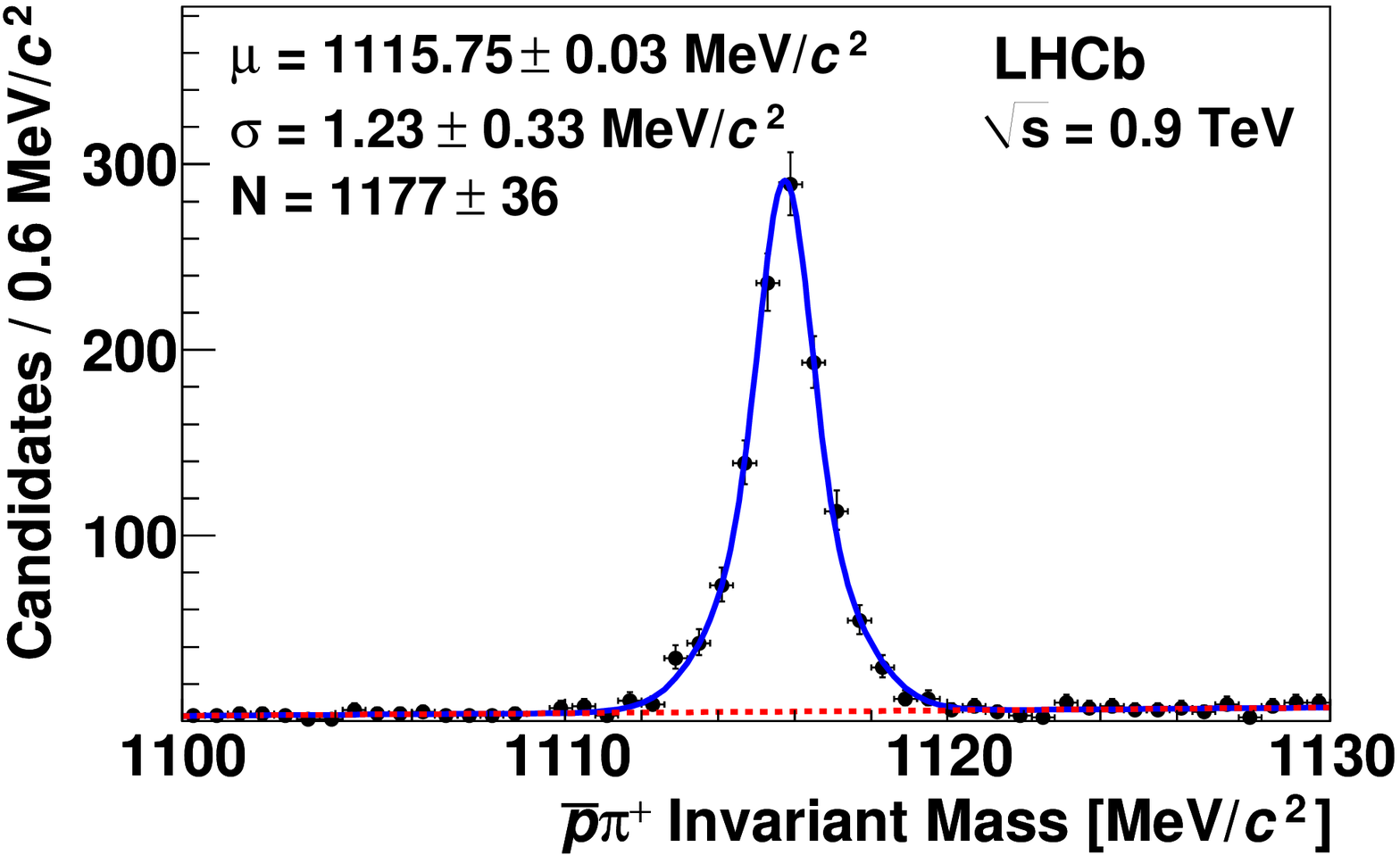}
    \label{fig:mass-fit-a}
}
\subfigure[]{
  \includegraphics[width=0.47\textwidth]{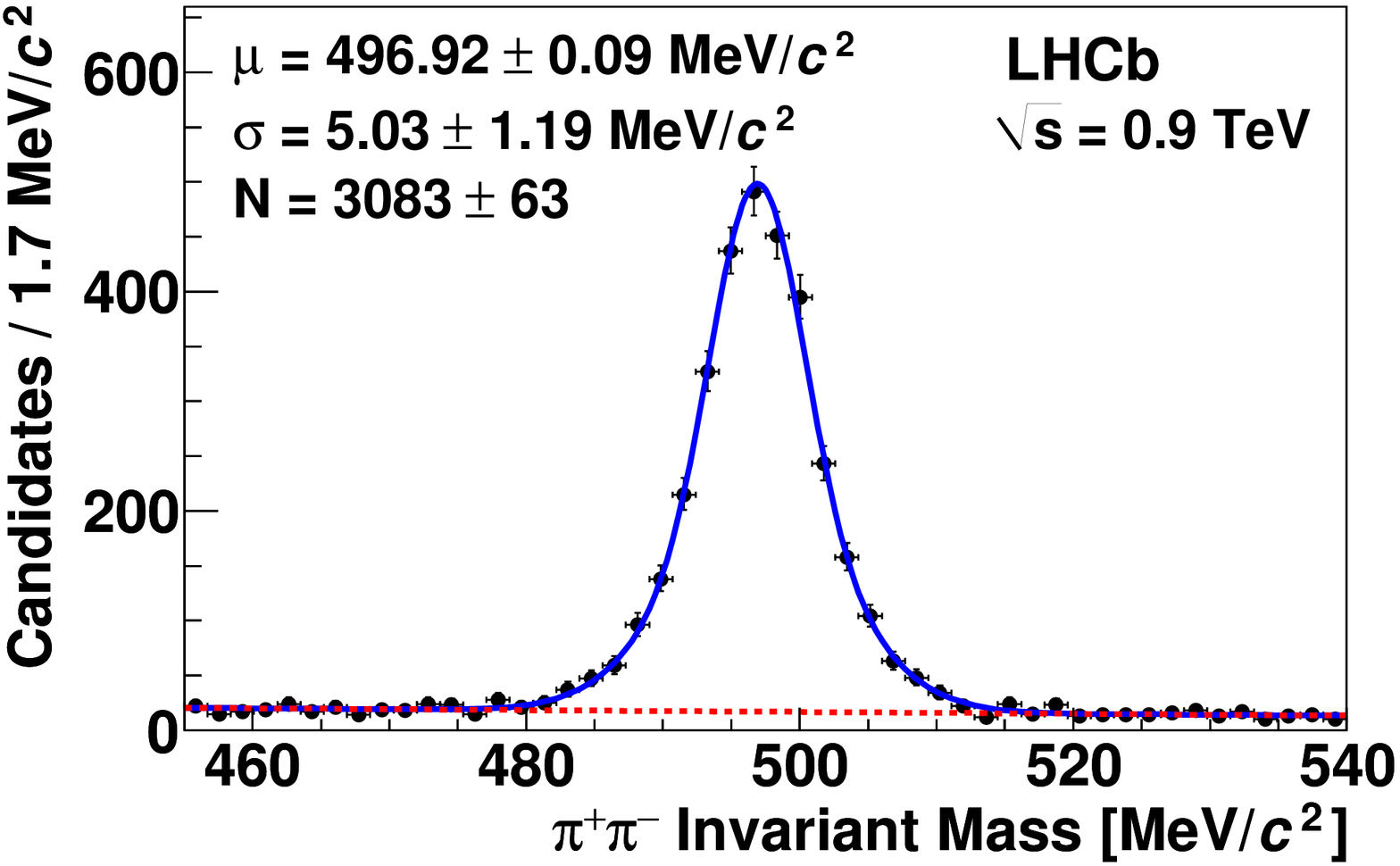}
    \label{fig:mass-fit-k}
}
\caption{\small Invariant mass peaks for \subref{fig:mass-fit-a} \Lbar in the range $0.25<\pt<2.50\gevc$ \& $2.5<\y<3.0$ and \subref{fig:mass-fit-k} \KS in the range $0.65<\pt<1.00\gevc$ \& $3.5<\y<4.0$ at \LHCLow with field \LHCbUp.  Signal yields, $N$, are found from fits (solid curves) with a double Gaussian peak with common mean, $\mu$, over a linear background (dashed lines).  The width, $\sigma$, is computed as the quadratic average of the two Gaussian widths weighted by their signal fractions.}
\label{fig:mass-fits}
\end{figure}

After the above selection, \VZ yields are estimated from data and simulation by fits to the invariant mass distributions, examples of which are shown in Fig.~\ref{fig:mass-fits}.  These fits are carried out with the method of unbinned extended maximum likelihood and are parametrised by a double Gaussian signal peak (with a common mean) over a linear background.  The mean values show a small, but statistically significant, deviation from the known \KS and \L\!(\Lbar) masses\,\cite{PDG}, reflecting the status of the momentum-scale calibration of the experiment.  The width of the peak is computed as the quadratic average of the two Gaussian widths, weighted by their signal fractions.  This width is found to be constant as a function of \pt and increases linearly toward higher \y, \eg by 1.4\,(0.8)\mevcc per unit rapidity for \KS\!(\L and \Lbar) at \LHCHigh.  The resulting signal yields are listed in Table~\ref{tab:signal-yields}.

\begin{figure}
 \centering
 \subfigure[]{
   \includegraphics[width=0.47\textwidth]{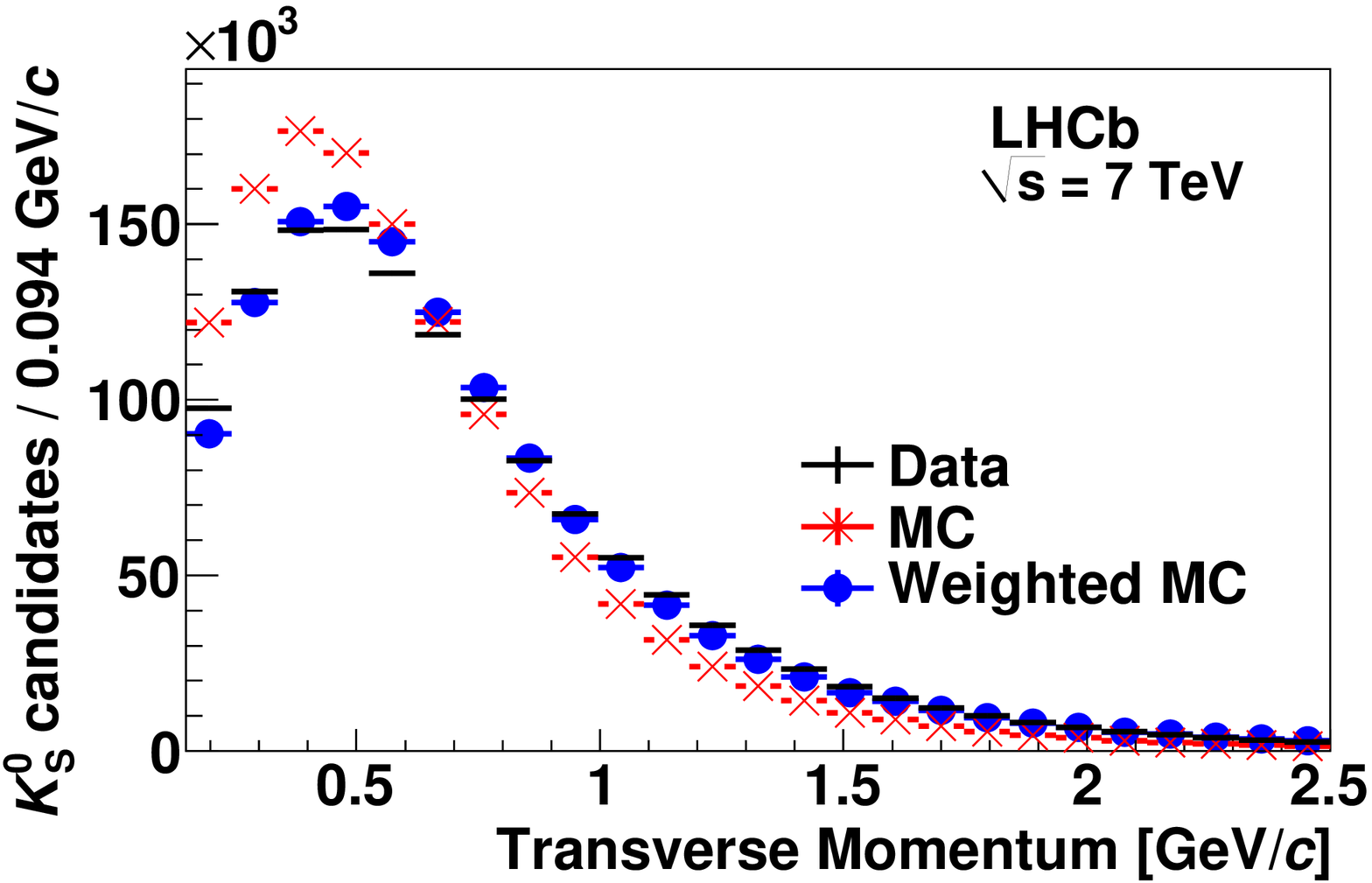}
     \label{fig:weighting-pt}
 }
 \subfigure[]{
   \includegraphics[width=0.47\textwidth]{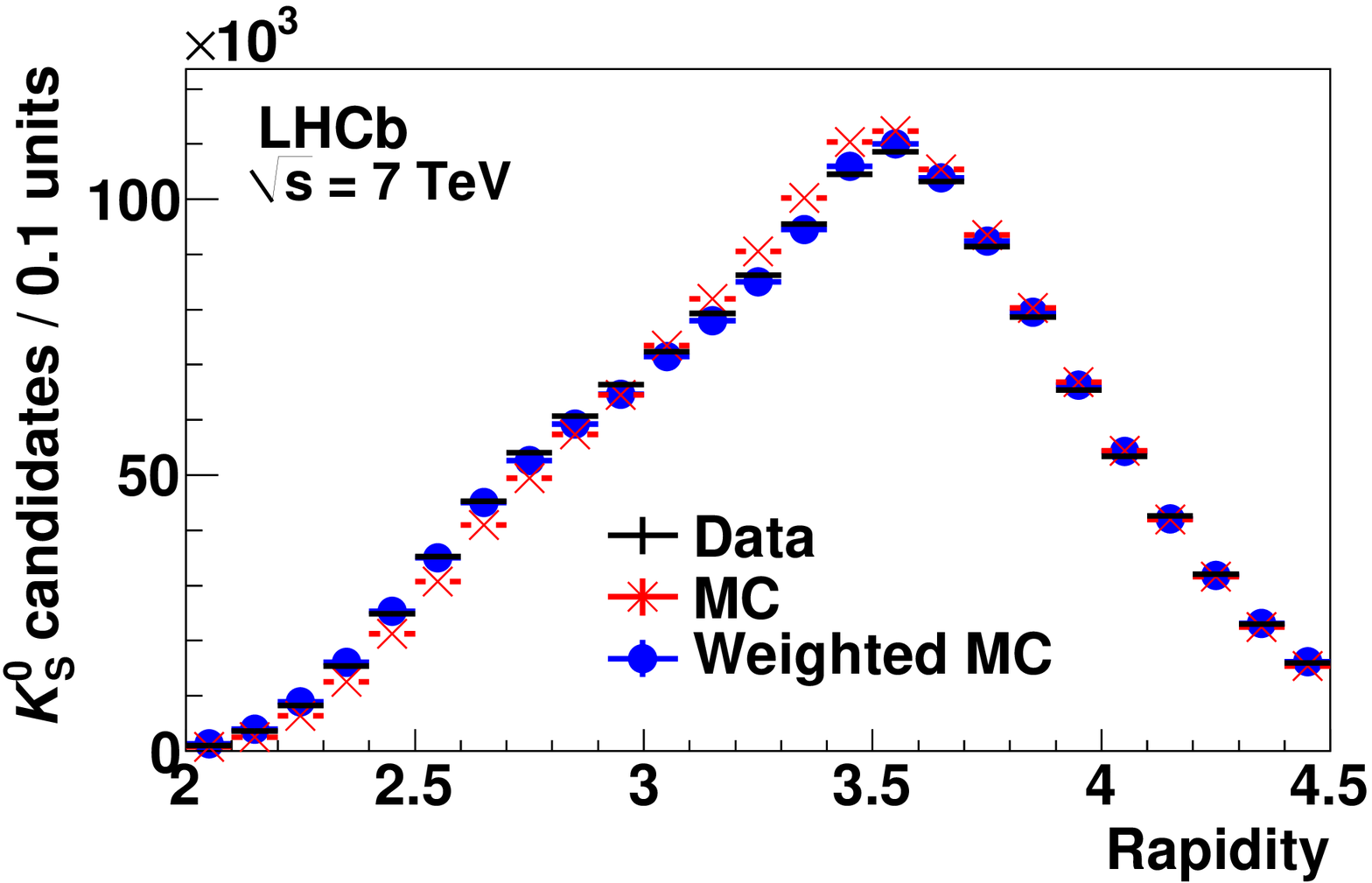}
     \label{fig:weighting-y}
 }
 \caption{\small \subref{fig:weighting-pt} Transverse momentum and \subref{fig:weighting-y} rapidity distributions for \KS in data and Monte Carlo simulation at \LHCHigh.  The difference between data and Monte Carlo is reduced by weighting the simulated candidates.}
 \label{fig:weighting}
\end{figure}

Significant differences are observed between \VZ kinematic variables reconstructed in data and in the simulation used for efficiency determination.  These differences can produce a bias for the measurement of \Lbar/\KS given the different production kinematics of the baryon and meson.  Simulated \VZ candidates are therefore weighted to match the two-dimensional \pt, \y distributions observed in data. These distributions are shown projected along both axes in Fig.~\ref{fig:weighting}.  The \VZ signal yield \pt, \y distributions are estimated from selected data and Monte Carlo candidates using sideband subtraction.  Two-dimensional fits, linear in both \pt and \y, are made to the ratios data/MC of these yields independently for \Lbar, \L and \KS, for each magnet polarity and collision energy.  The resulting functions are used to weight generated and selected \VZ candidates in the Monte Carlo simulation.  These weights vary across the measured \pt, \y range between 0.4 and 2.1, with typical values between 0.8 and 1.2.  

\begin{table}
\centering
\caption{\small Integrated signal yields extracted by fits to the invariant mass distributions of selected \VZ candidates from data taken with magnetic field \LHCbUp and \LHCbDown at $\s = 0.9$ and 7\tev.}
\hfill\\
\small
\begin{tabular}{c|cc|cc}
\toprule
\s & \multicolumn{2}{c|}{0.9\tev} & \multicolumn{2}{c}{7\tev} \\
Magnetic field  & Up & Down & Up & Down \\
\midrule
\Lbar  &  $3,440\pm60$  &  $4,100\pm70$  & $258,930\pm640$ & $132,550\pm460$ \\
\L     &  $4,880\pm80$  &  $5,420\pm80$  & $294,010\pm680$ & $141,860\pm460$ \\
\KS    & $35,790\pm200$ & $40,230\pm220$ & $2,737,090\pm1,940$ & $1,365,990\pm1,370$ \\
%\Lbar  &  $3,442\pm64$  &  $4,096\pm72$  & $258,927\pm642$ & $132,548\pm459$ \\
%\L     &  $4,877\pm75$  &  $5,416\pm80$  & $294,005\pm677$ & $141,861\pm463$ \\
%\KS    & $35,785\pm203$ & $40,234\pm219$ & $2,737,093\pm1,935$ & $1,365,993\pm1,365$ \\
\bottomrule
\end{tabular}
\label{tab:signal-yields}
\end{table}

The measured ratios are presented in three complementary binning schemes: projections over the full \pt range, the full \y range, and a coarser two-dimensional binning.  The rapidity range \mbox{$2.0 < \y < 4.0$\,(4.5)} is split into 0.5-unit bins, while six bins in \pt are chosen to approximately equalise signal \VZ statistics in data over the range \mbox{0.25\,$(0.15) < \pt < 2.50$}\gevc from collisions at $\s = 0.9$\,(7)\tev.  The two-dimensional binning combines pairs of \pt bins. The full analysis procedure is carried out independently in each \pt, \y bin.

The efficiency for selecting prompt \VZ decays is estimated from simulation as
\begin{equation} \label{equ:efficiency}
  \varepsilon = \frac{ N(\VZ \to d^{+}d^{-})_{\mathrm{Observed}} }{ N(pp \to \VZ X)_{\mathrm{Generated}} },
\end{equation}
\noindent where the denominator is the number of prompt \VZ hadrons generated in a given \pt, \y region after weighting and the numerator is the number of those weighted candidates found from the selection and fitting procedure described above.  The efficiency therefore accounts for decays via other channels and losses from interactions with the detector material.  Prompt \VZ hadrons are defined in Monte Carlo simulation by the cumulative lifetimes of their ancestors
\begin{equation} \label{equ:prompt-cut}
\sum\limits_{i=1}^n c\tau_{i} < 10^{-9}\m,
\end{equation}
\noindent where $\tau_{i}$ is the proper decay time of the $i^{\textrm{th}}$ ancestor.  This veto is defined such as to keep only \VZ hadrons created either directly from the \pp collisions or from the strong or electromagnetic decays of particles produced at those collisions, removing \VZ hadrons generated from material interactions and weak decays.  The Fisher discriminant \FIP strongly favours prompt \VZ hadrons, however a small non-prompt contamination in data would lead to a systematic bias in the ratios.  The fractional contamination of selected events is determined from simulation to be $2-6\,\%$ for \Lbar and \L, depending on the measurement bin, and about 1\,\% for \KS.  This effect is dominated by weak decays rather than material interactions.  The resulting absolute corrections to the ratios \Lbar/\L and \Lbar/\KS are approximately 0.01.

%% -----------------------------------------------------------------
%% Measurement of V0 production ratios in pp collisions at _/s = 0.9 and 7 TeV
%% CERN-PH-EP-2011-082
%% -----------------------------------------------------------------
%% LHCb Collaboration 
%% T. Blake, C. Blanks, W. Bonivento, F. Dettori, R. Muresan
%% c.blanks07@imperial.ac.uk
%% -----------------------------------------------------------------

\section{Systematic uncertainties}\label{sec:systematics}

The measured efficiency corrected ratios \Lbar/\L and \Lbar/\KS are subsequently corrected for non-prompt contamination as found from Monte Carlo simulation and defined by Eq.~\ref{equ:prompt-cut}.  This procedure relies on simulation and the corrections may be biased by the choice of the \lhcb MC generator tune.  To estimate a systematic uncertainty on the correction for non-prompt \VZ, the contaminant fractions are also calculated using two alternative tunes of \pythia\!6: Perugia\,0 and Perugia\,NOCR\,\cite{Peter}.  The maximum differences in non-prompt fraction across the measurement range and at both energies are $<1\,\%$ for each \VZ species.  The resulting absolute uncertainties on the ratios are $<0.01$.

The efficiency of primary vertex reconstruction may introduce a bias on the measured ratios if the detector occupancy is different for events containing \KS, \L or \Lbar.  This efficiency is compared in data and simulation using \VZ samples obtained with an alternative selection not requiring a primary vertex.  Instead, the \VZ flight vector is extrapolated towards the beam axis to find the point of closest approach.  The $z$ coordinate of this point is used to define a pseudo-vertex, with $x=y=0$.  Candidates are kept if the impact parameters of their daughter tracks to this pseudo-vertex are $>0.2$\,mm.  There is a large overlap of signal candidates with the standard selection.  The primary vertex finding efficiency is then explored by taking the ratio of these selected events which do or do not have a standard primary vertex.  Calculated in bins of \pt and \y, this efficiency agrees between data and simulation to better than 2\,\% at both $\s = 0.9$ and 7\tev. The resulting absolute uncertainties on \Lbar/\L and \Lbar/\KS are $<0.02$ and $<0.01$, respectively.
 
\begin{figure}
\subfigure[]{
\includegraphics[width = 0.48\textwidth]{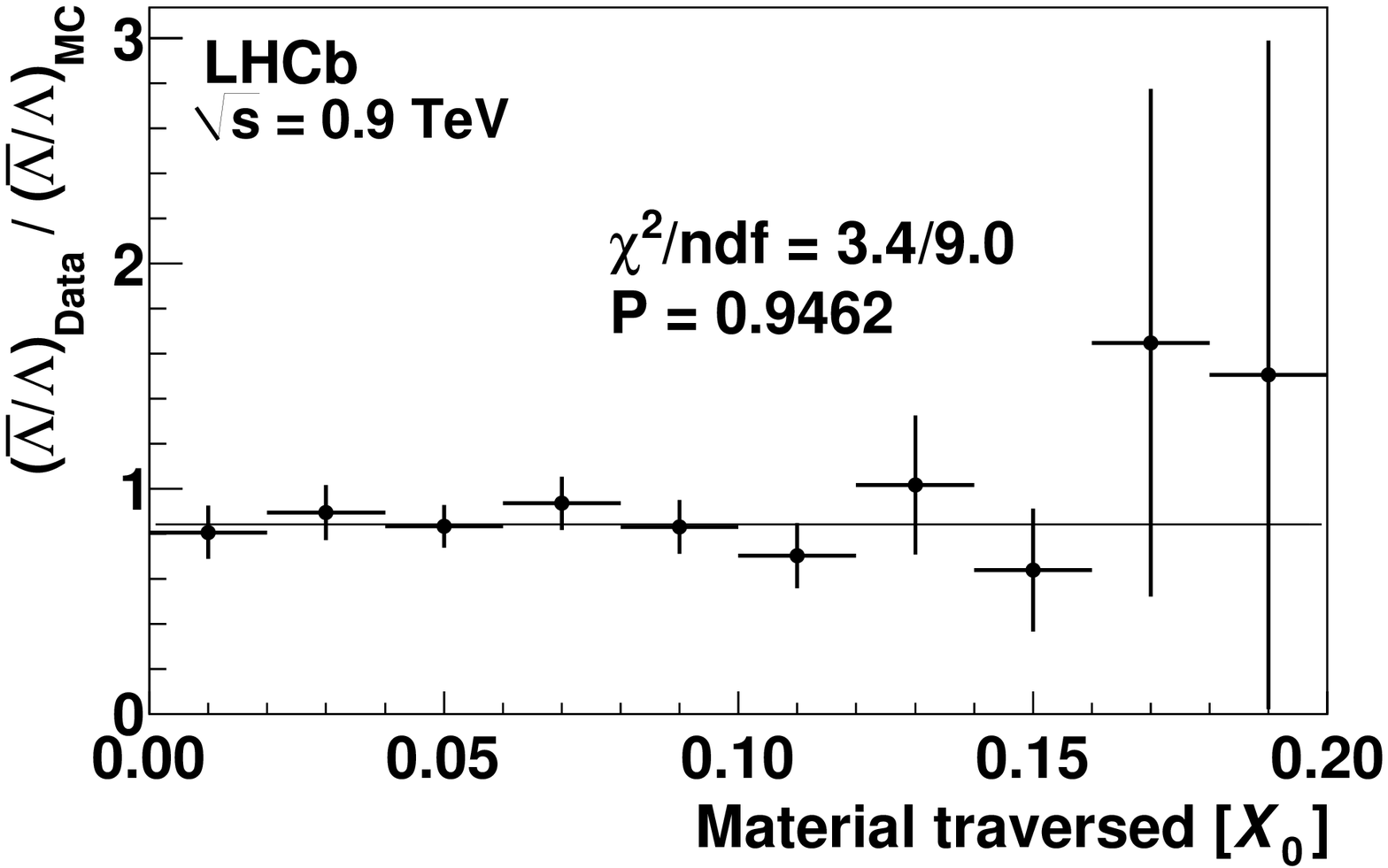}
\label{fig:material-int-al}
}
\subfigure[]{
\includegraphics[width = 0.48\textwidth]{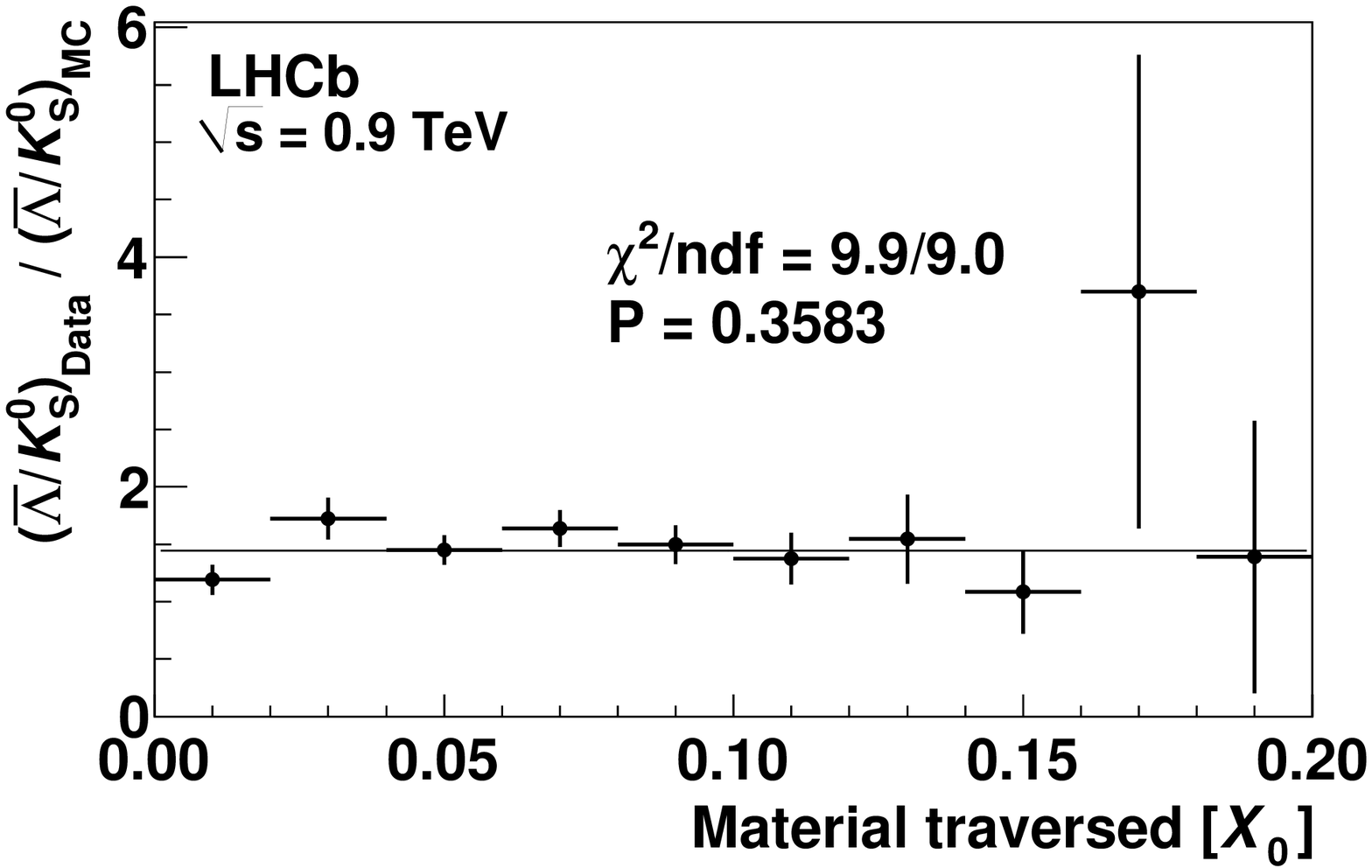}
\label{fig:material-int-ak}
}
\caption{\small The double ratios \subref{fig:material-int-al} $(\Lbar/\L)_\mathrm{Data}/(\Lbar/\L)_\mathrm{MC}$ and \subref{fig:material-int-ak}  $(\Lbar/\KS)_\mathrm{Data}/(\Lbar/\KS)_\mathrm{MC}$ are shown as a function of the material traversed, in units of radiation length.  Flat line fits, shown together with their respective $\chi^2$ probabilities, give no evidence of a bias.}
\label{fig:material-int-ratios}
\end{figure}

The primary vertex finding algorithm requires at least three reconstructed tracks.\footnote{The minimum requirements for primary vertex reconstruction at \lhcb can be approximated in Monte Carlo simulation by a generator-level cut requiring at least three charged particles from the collision with lifetime $c\tau > 10^{-9}\m$, momentum $p>0.3\gevc$ and polar angle $15<\theta<460\mrad$.}  Therefore, the reconstruction highly favours non-diffractive events due to the relatively low efficiency for finding diffractive interaction vertices, which tend to produce fewer tracks.  In the \lhcb MC simulation, the diffractive cross-section accounts for 28\,(25)\,\% of the total minimum-bias cross-section of 65\,(91)\,mb at 0.9\,(7)\tev\!\cite{Gauss}.  Due to the primary vertex requirement, only about 3\,\% of the \VZ candidates selected in simulation are produced in diffractive events.  These fractions are determined using \pythia~6 which models only soft diffraction.  As a cross check, the fractions are also calculated with \pythia~8 which includes both soft and hard diffraction.  The variation on the overall efficiency between models is about 2\,\% for both ratios at \LHCHigh and close to 1\,\% at 0.9\tev.  Indeed, complete removal of diffractive events only produces a change of $0.01-0.02$ in the ratios across the measurement range.

The track reconstruction efficiency depends on particle momentum.  In particular, the tracking efficiency varies rapidly with momentum for tracks below 5\gevc.  Any bias is expected to be negligible for the ratio \Lbar/\L but can be larger for \Lbar/\KS due to the different kinematics.  Two complementary procedures are employed to check this efficiency.  First, track segments are reconstructed in the tracking stations upstream of the magnet.  These track segments are then paired with the standard tracks reconstructed through the full detector and the pairs are required to form a \KS to ensure only genuine tracks are considered.  This track matching gives a measure of the tracking efficiency for the upstream tracking systems.  The second procedure uses the downstream stations to reconstruct track segments, which are similarly paired with standard tracks to measure the efficiency of the downstream tracking stations.   The agreement between these efficiencies in data and simulation is better than 5\,\%.  To estimate the resulting uncertainty on \Lbar/\L and \Lbar/\KS, both ratios are re-calculated after weighting \VZ candidates by $95\,\%$ for each daughter track with momentum below 5\gevc.  The resulting systematic shifts in the ratios are $<0.01$.

\begin{table}
 \centering
  \caption{\small Absolute systematic errors are listed in descending order of importance.  Ranges indicate uncertainties that vary across the measurement bins and/or by collision energy.  Correlated sources of uncertainty between field \LHCbUp and \LHCbDown are identified.}
\hfill\\
\small
 \begin{tabular}{lcc}
\toprule
 Sources of systematic uncertainty & \Lbar/\L & \Lbar/\KS \\
\midrule
\emph{Correlated between field \LHCbUp and \LHCbDown}: & & \\ 
Material interactions               & $0.02$ & $0.02$ \\ 
Diffractive event fraction          & $0.01-0.02$ & $0.01-0.02$ \\ 
Primary vertex finding              & $<0.02$ & $<0.01$ \\ 
Non-prompt fraction                 & $<0.01$  & $<0.01$  \\ 
Track finding                       & negligible & $0.01$ \\
\midrule
 \emph{Uncorrelated}: &  &  \\  
Kinematic correction             & $0.01-0.05$ & $<0.03$ \\ 
Signal extraction from fit         & $0.001$ & $0.001$ \\
\midrule
 Total & $0.02-0.06$ & $0.02-0.03$ \\ 
\bottomrule
 \end{tabular}
\label{tab:syst-summary}
\end{table}
 
Particle interactions within the detector are simulated using the \geant package, which implements interaction cross-sections for each particle according to the LHEP physics list\,\cite{Geant}.  These simulated cross-sections have been tested in the \lhcb framework and are consistent with the LHEP values.  The small measured differences are propagated to \Lbar/\L and \Lbar/\KS to estimate absolute uncertainties on the ratios of about 0.02.  \VZ absorption is limited by the requirement that each \VZ decay occurs within the most upstream tracker (the VELO).  Secondary \VZ production in material is suppressed by the Fisher discriminant, which rejects \VZ candidates with large impact parameter.  The potential bias on the ratios is explored by measurement of both \Lbar/\L and \Lbar/\KS as a function of material traversed (determined by the detector simulation), in units of radiation length, $X_0$.   Data and simulation are compared by their ratio, shown in Fig.~\ref{fig:material-int-ratios}.  These double ratios are consistent with a flat line as a function of $X_0$, therefore any possible imperfections in the description of the detector material in simulation do not have a large effect on the \VZ ratios. Note that the double ratios are not expected to be unity since simulations do not predict the same values for \Lbar/\L and \Lbar/\KS as are observed in data.

The potential bias from the Fisher discriminant, \FIP, is investigated using a pre-selected sample, with only the track and vertex quality cuts applied.  The distributions of \FIP for \L, \Lbar and \KS in data and Monte Carlo simulation are estimated using sideband subtraction.  The double ratios of data/MC efficiencies are seen to be independent of the discriminant, implying that the distribution is well modelled in the simulation.  No systematic uncertainty is assigned to this selection requirement.

A degradation is observed of the reconstructed impact parameter resolution in data compared to simulation.  The simulated \VZ impact parameters are recalculated with smeared primary and secondary vertex positions to match the resolution measured in data.  There is a negligible effect on the \VZ ratio results.

A good estimate of the reconstructed yields and their uncertainties in both data and simulation is provided by the fitting procedure but there may be a residual systematic uncertainty from the choice of this method.  Comparisons are made using side-band subtraction and the resulting \VZ yields are in agreement with the results of the fits at the 0.1\,\% level.  The resulting absolute uncertainties on the ratios are on the order of 0.001.

Simulated events are weighted to improve agreement between simulated \VZ kinematic distributions and data.  As described in Section~\ref{sec:analysis}, these weights are calculated from a two-dimensional fit, linear in both \pt and \y, to the distribution of the ratio between reconstructed data and simulated Monte Carlo candidates.  This choice of parametrisation could be a source of systematic uncertainty, therefore alternative procedures are investigated including a two-dimensional polynomial fit to $3^{\textrm{rd}}$ order in both \pt and \y and a (non-parametric) bilinear interpolation.  The results from each method are compared across the measurement range to estimate typical systematic uncertainties of $0.01-0.05$ for \Lbar/\L and $<0.03$ for \Lbar/\KS.

The lifetime distributions of reconstructed and selected \VZ candidates are consistent between data and simulation.  The possible influence of transverse \L (\Lbar) polarisation was explored by simulations with extreme values of polarisation and found to produce no significant effect on the measured ratios. Potential acceptance effects were checked as a function of azimuthal angle, with no evidence of systematic bias.  The potential sources of systematic uncertainty or bias are summarised in Table~\ref{tab:syst-summary}. 

%% -----------------------------------------------------------------
%% Measurement of V0 production ratios in pp collisions at _/s = 0.9 and 7 TeV
%% CERN-PH-EP-2011-082
%% -----------------------------------------------------------------
%% LHCb Collaboration 
%% T. Blake, C. Blanks, W. Bonivento, F. Dettori, R. Muresan
%% c.blanks07@imperial.ac.uk
%% -----------------------------------------------------------------

\begin{figure}[t!]
  \centering
  \subfigure[]{
    \includegraphics[width=0.47\textwidth]{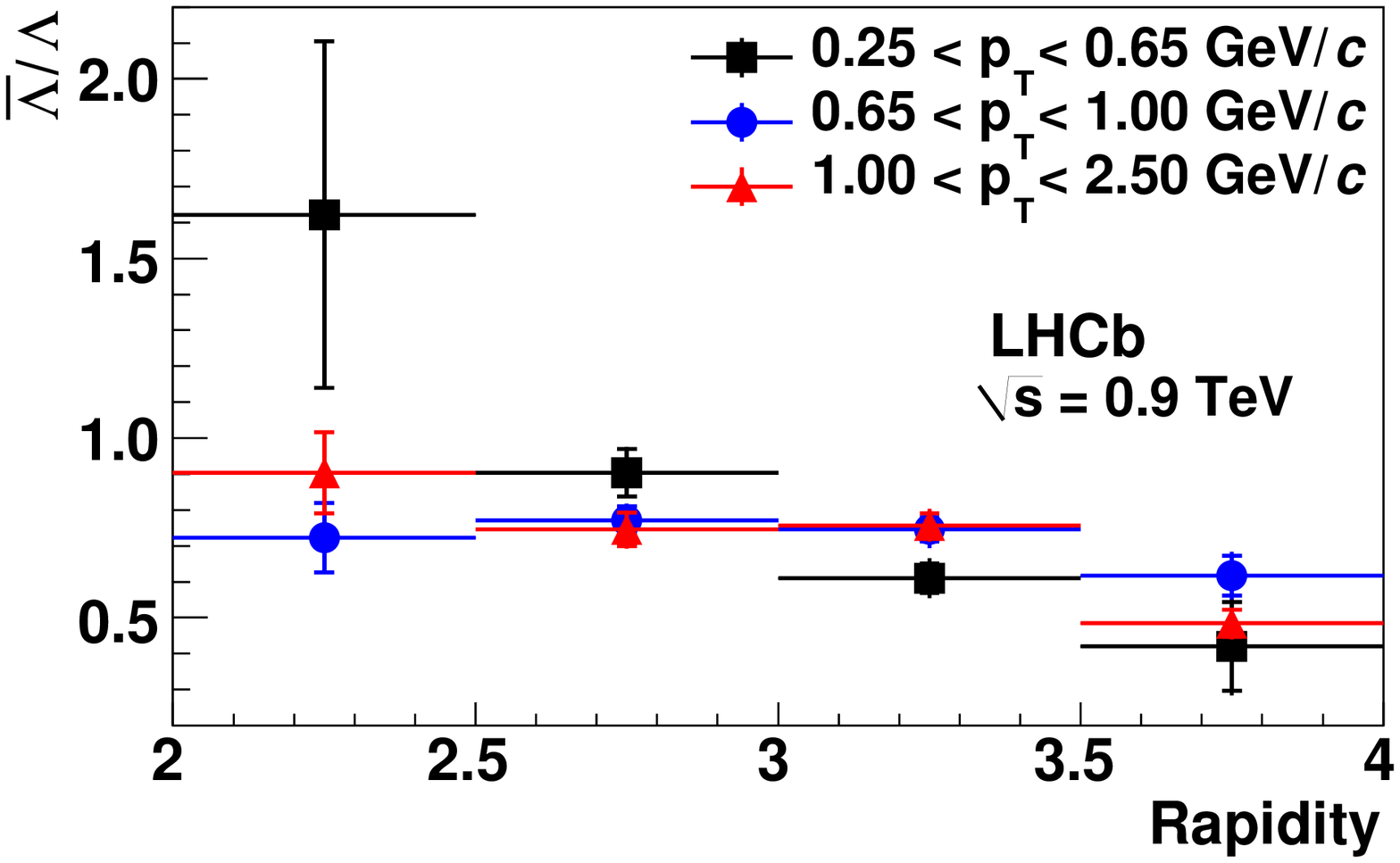}
    \label{fig:plotData-900-al-3y}
  }
  \subfigure[]{
    \includegraphics[width=0.47\textwidth]{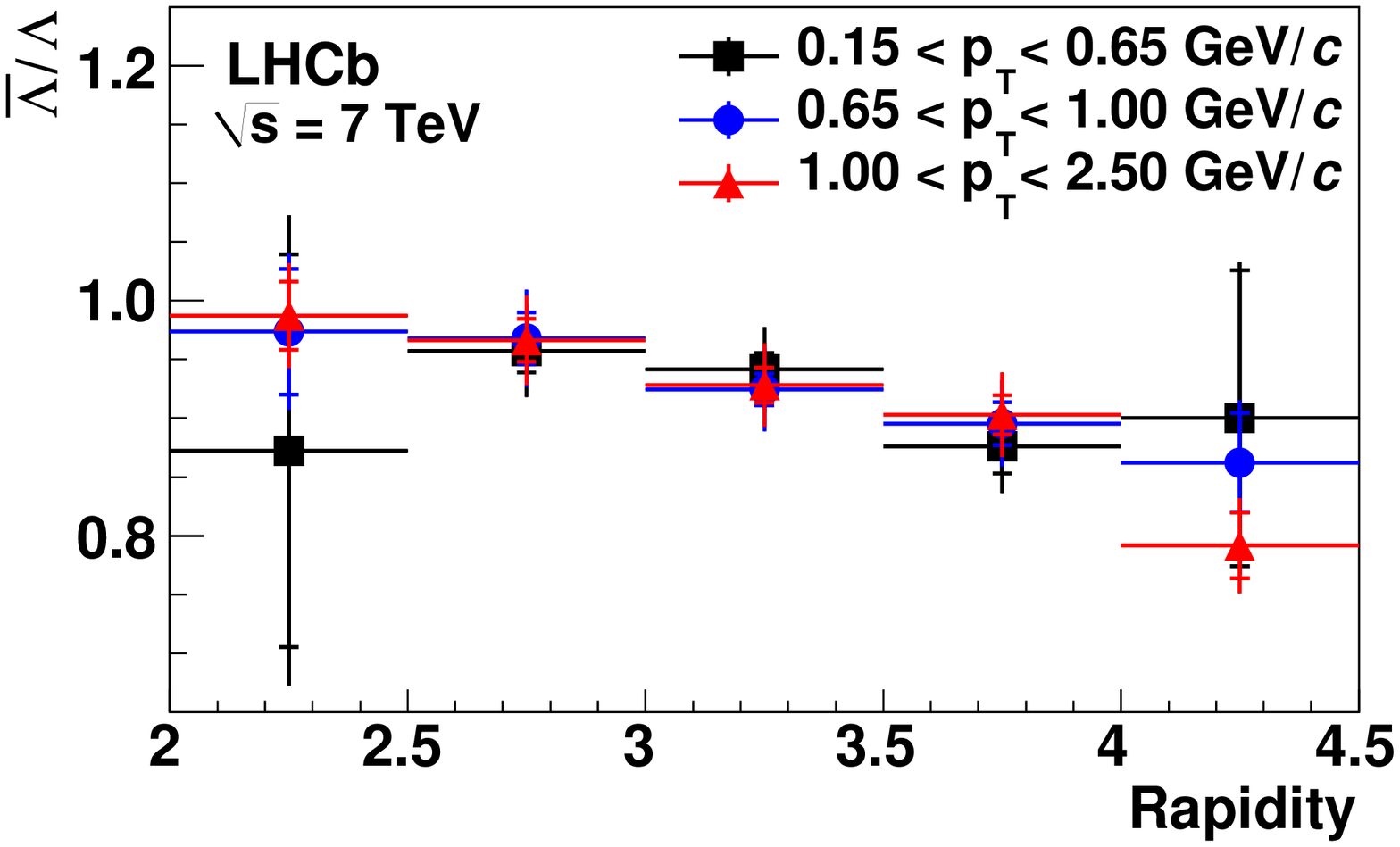}
    \label{fig:plotData-7-al-3y}
  }
  \subfigure[]{
    \includegraphics[width=0.47\textwidth]{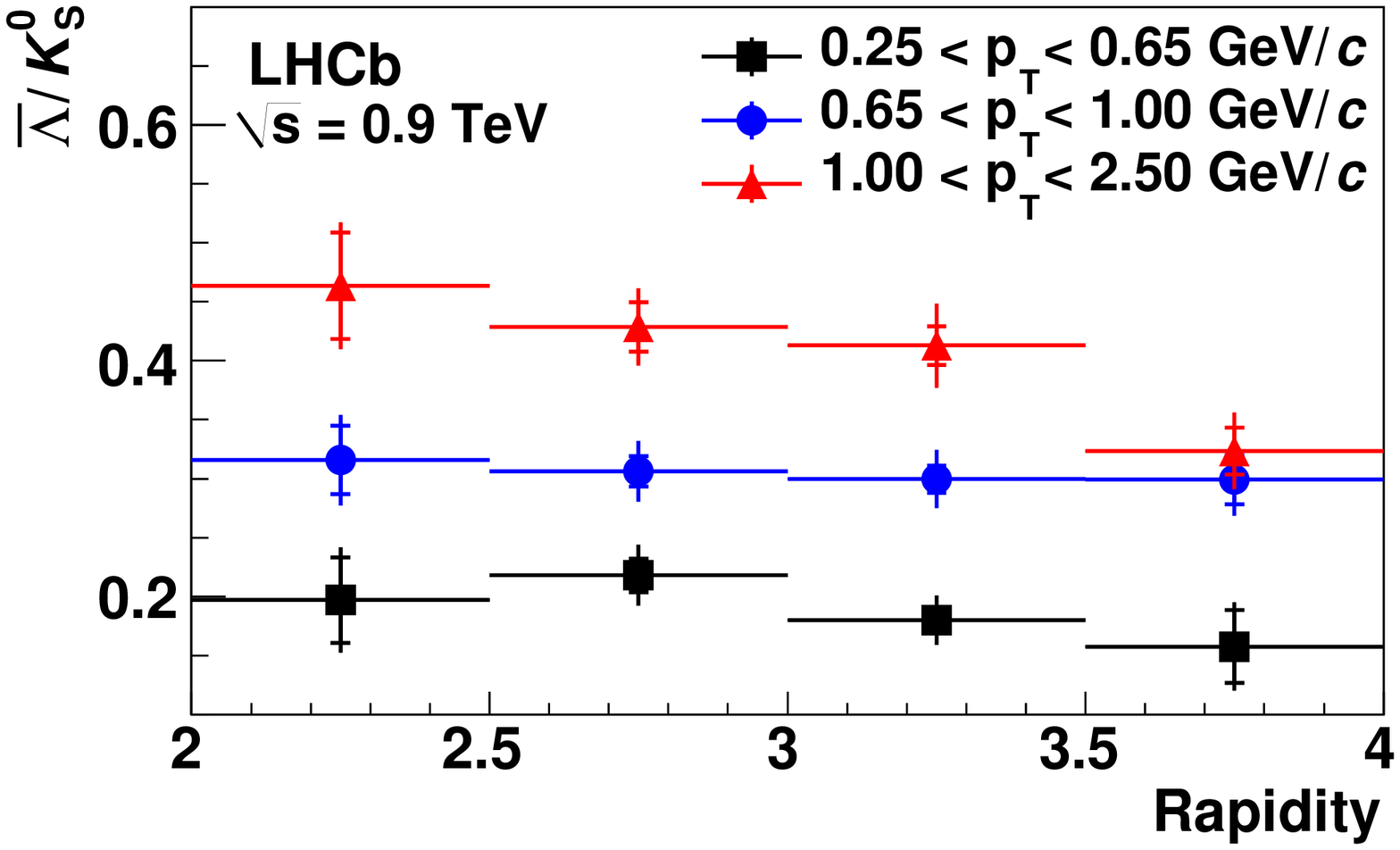}
    \label{fig:plotData-900-ak-3y}
  }
  \subfigure[]{
    \includegraphics[width=0.47\textwidth]{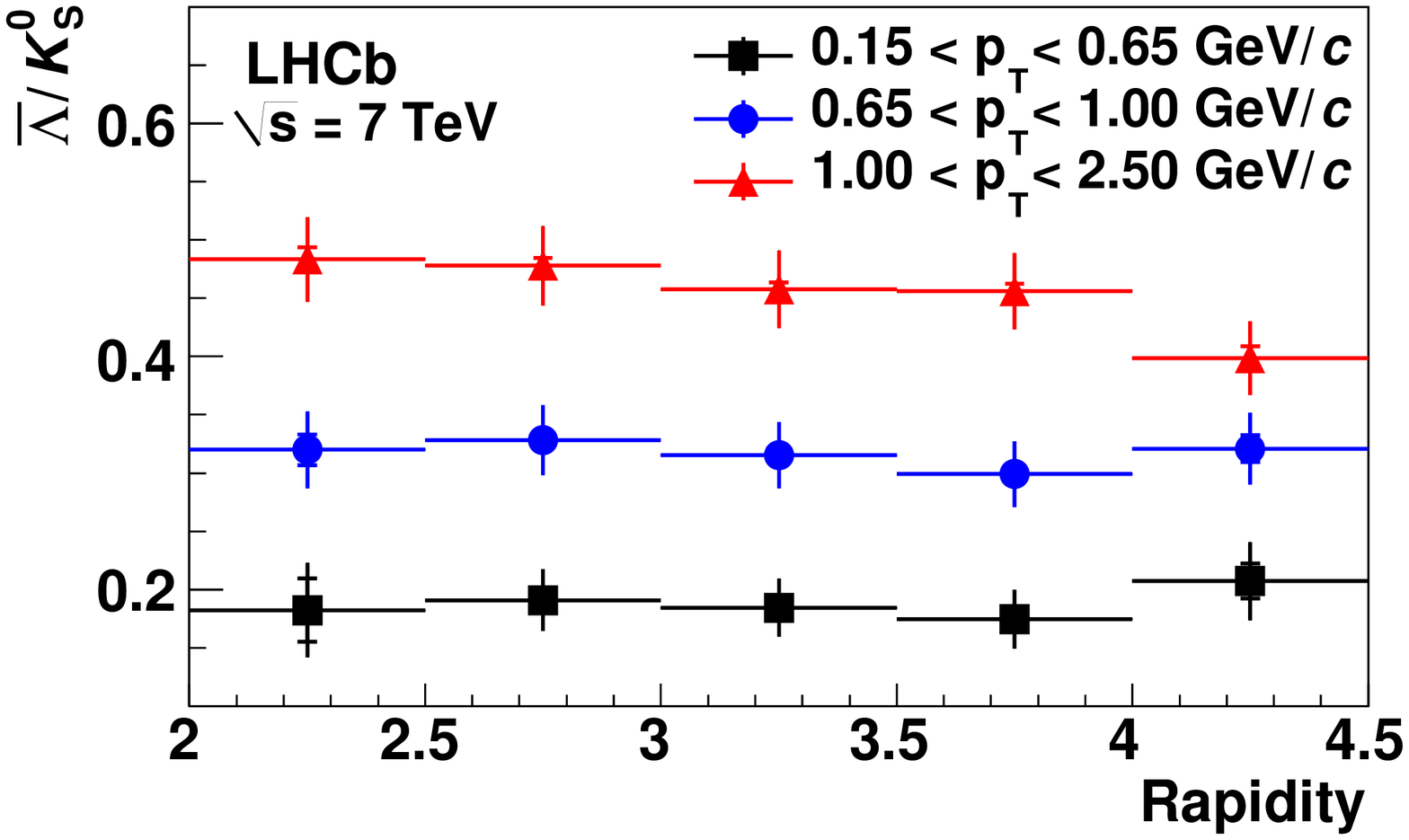}
    \label{fig:plotData-7-ak-3y}
  }
  \caption{\small The ratios \Lbar/\L and \Lbar/\KS from the full analysis procedure at \subref{fig:plotData-900-al-3y} \& \subref{fig:plotData-900-ak-3y} \LHCLow and \subref{fig:plotData-7-al-3y} \& \subref{fig:plotData-7-ak-3y} 7\tev are shown as a function of rapidity, compared across intervals of transverse momentum.  Vertical lines show the combined statistical and systematic uncertainties and the short horizontal bars (where visible) show the statistical component.}
  \label{fig:plotData3y}
\end{figure}

\section{Results}
\label{sec:results}

The \Lbar/\L and \Lbar/\KS production ratios are measured independently for each magnetic field polarity.  These measurements show good consistency after correction for detector acceptance.  Bin-by-bin comparisons in the two-dimensional binning scheme give $\chi^2$ probabilities for \Lbar/\L\!(\Lbar/\KS) of 3\,(18)\,\% at \LHCLow and 19\,(97)\,\% at \LHCHigh, with 12\,(15) degrees of freedom. The field \LHCbUp and \LHCbDown results are therefore combined to maximise statistical significance.  A weighted average is computed such that the result has minimal variance while taking into account the correlations between sources of systematic uncertainty identified in Table~\ref{tab:syst-summary}.  These combined results are shown as a function of \y in three intervals of \pt in Fig.~\ref{fig:plotData3y} at \LHCLow and 7\tev. The ratio \Lbar/\KS shows a strong \pt dependence.

\begin{figure}[t!]
  \centering
  \subfigure[]{
    \includegraphics[width=0.47\textwidth]{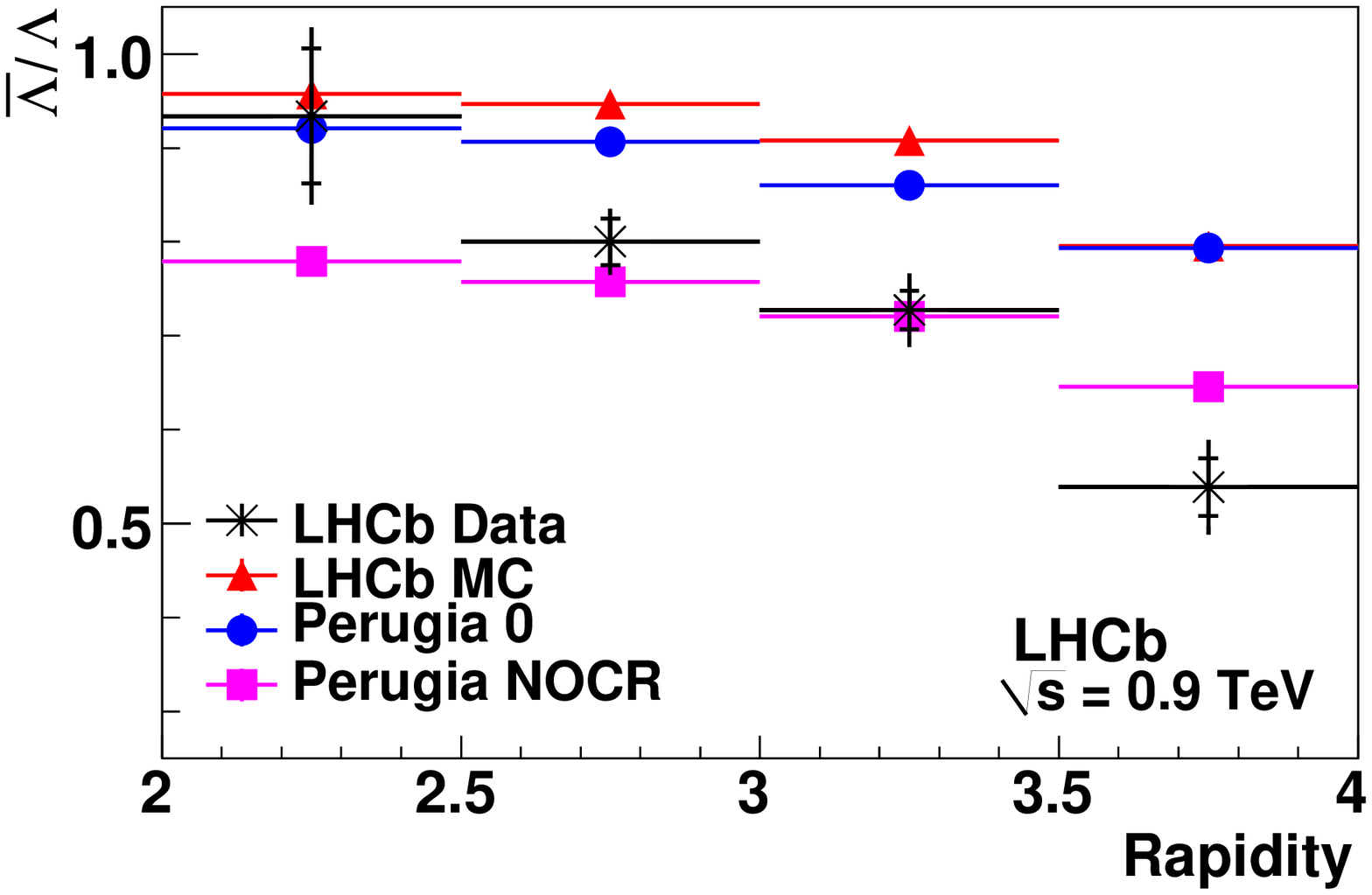}
    \label{fig:plotDataTheory-900-al-y}
  }
  \subfigure[]{
    \includegraphics[width=0.47\textwidth]{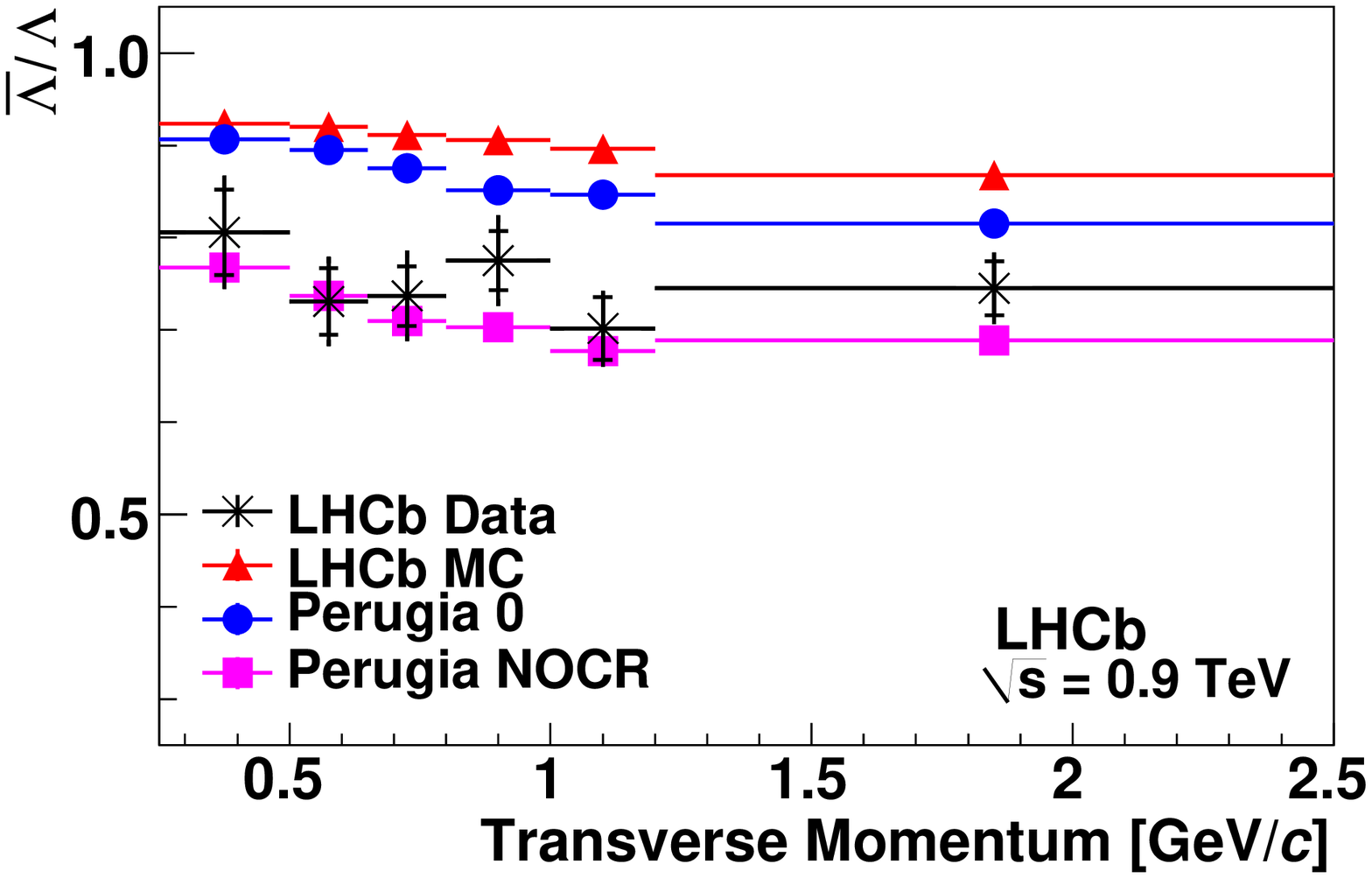}
    \label{fig:plotDataTheory-900-al-pt}
  }
  \subfigure[]{
    \includegraphics[width=0.47\textwidth]{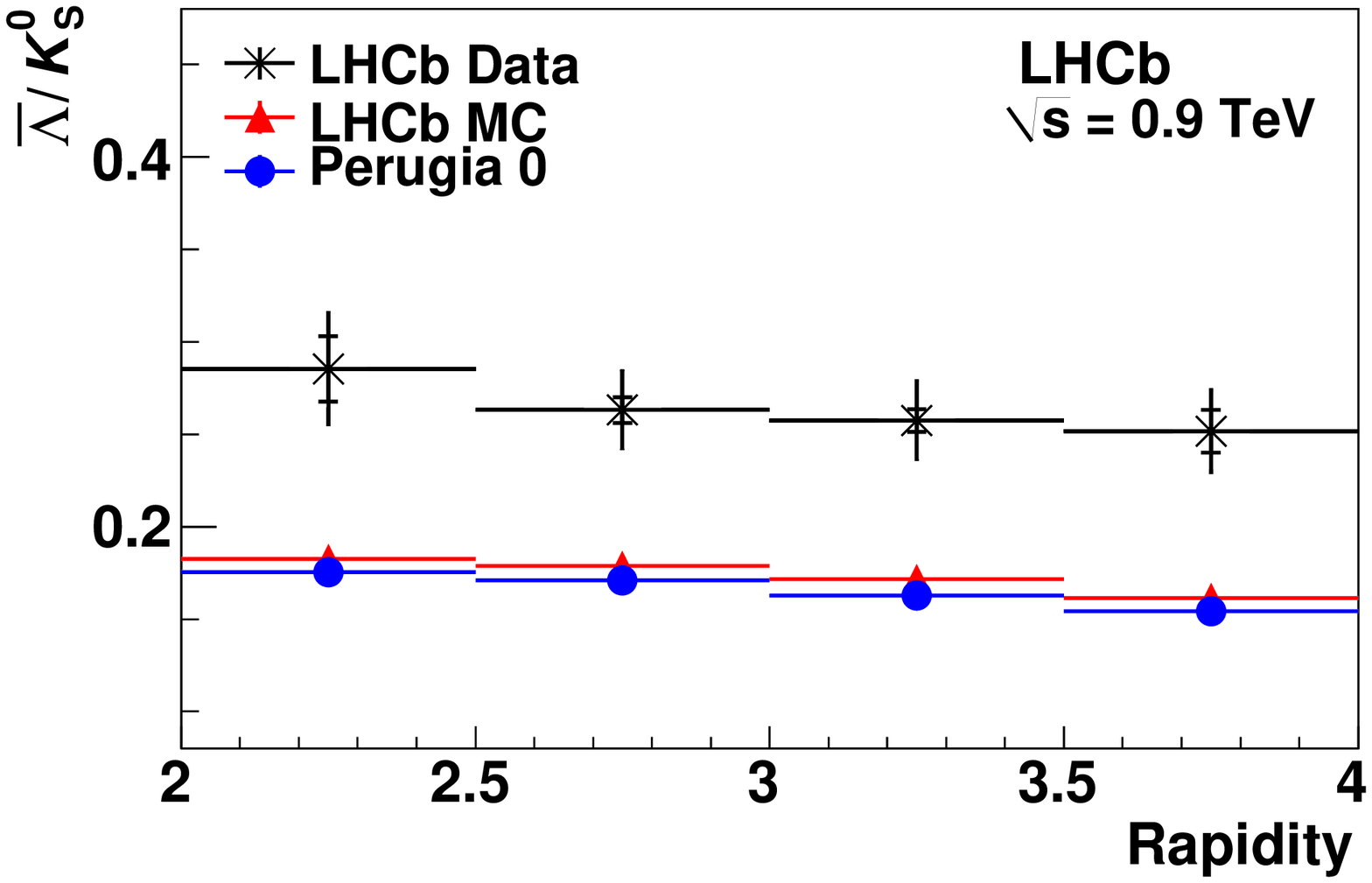}
    \label{fig:plotDataTheory-900-ak-y}
  }
  \subfigure[]{
    \includegraphics[width=0.47\textwidth]{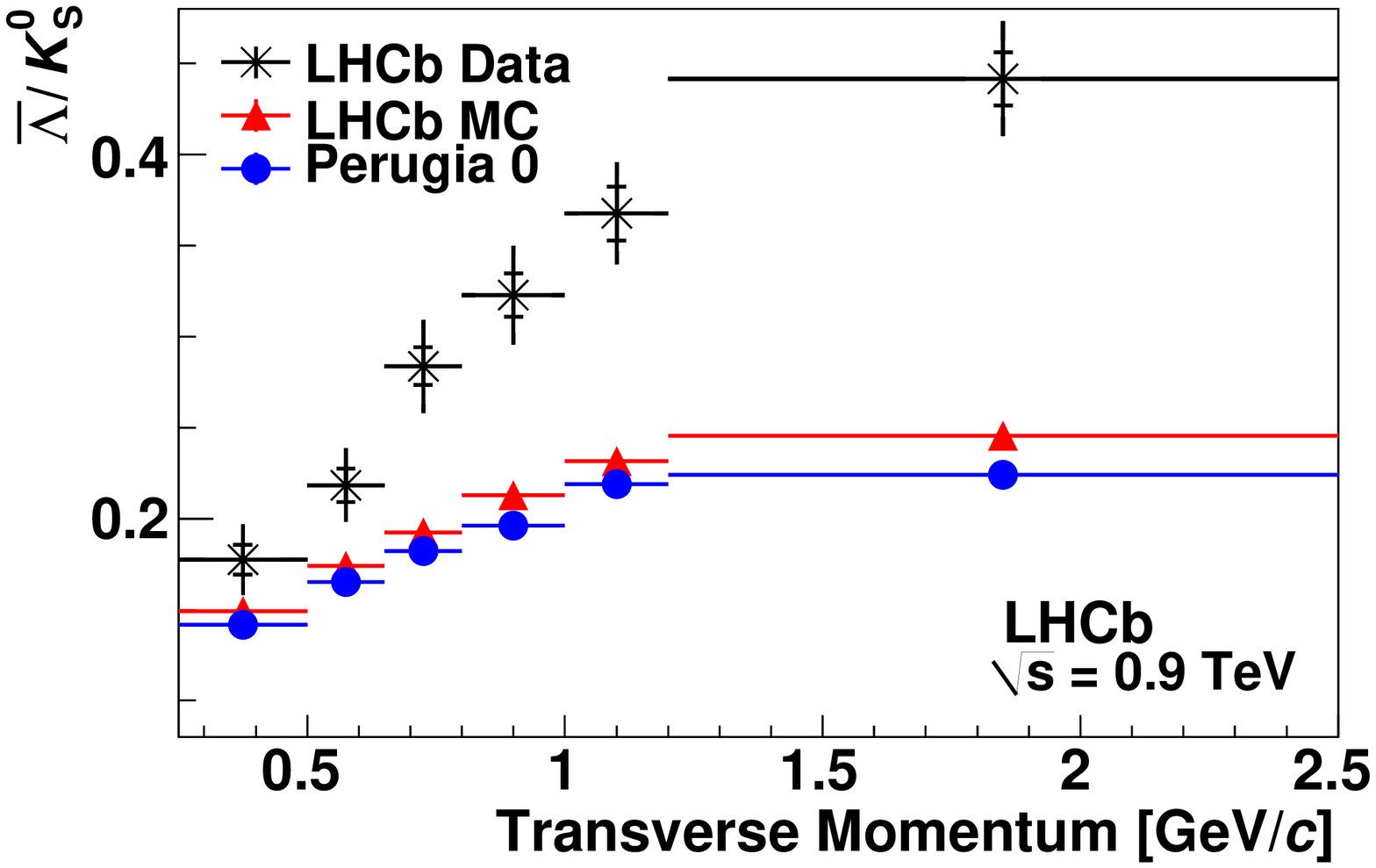}
    \label{fig:plotDataTheory-900-ak-pt}
  }
  \caption{\small The ratios \Lbar/\L and \Lbar/\KS at \LHCLow are compared with the predictions of the LHCb\,MC, Perugia\,0 and Perugia\,NOCR as a function of \subref{fig:plotDataTheory-900-al-y} \& \subref{fig:plotDataTheory-900-ak-y} rapidity and \subref{fig:plotDataTheory-900-al-pt} \& \subref{fig:plotDataTheory-900-ak-pt} transverse momentum.  Vertical lines show the combined statistical and systematic uncertainties and the short horizontal bars (where visible) show the statistical component.}
  \label{fig:plotDataTheory-900}
\end{figure}

\begin{figure}
  \centering
  \subfigure[]{
    \includegraphics[width=0.47\textwidth]{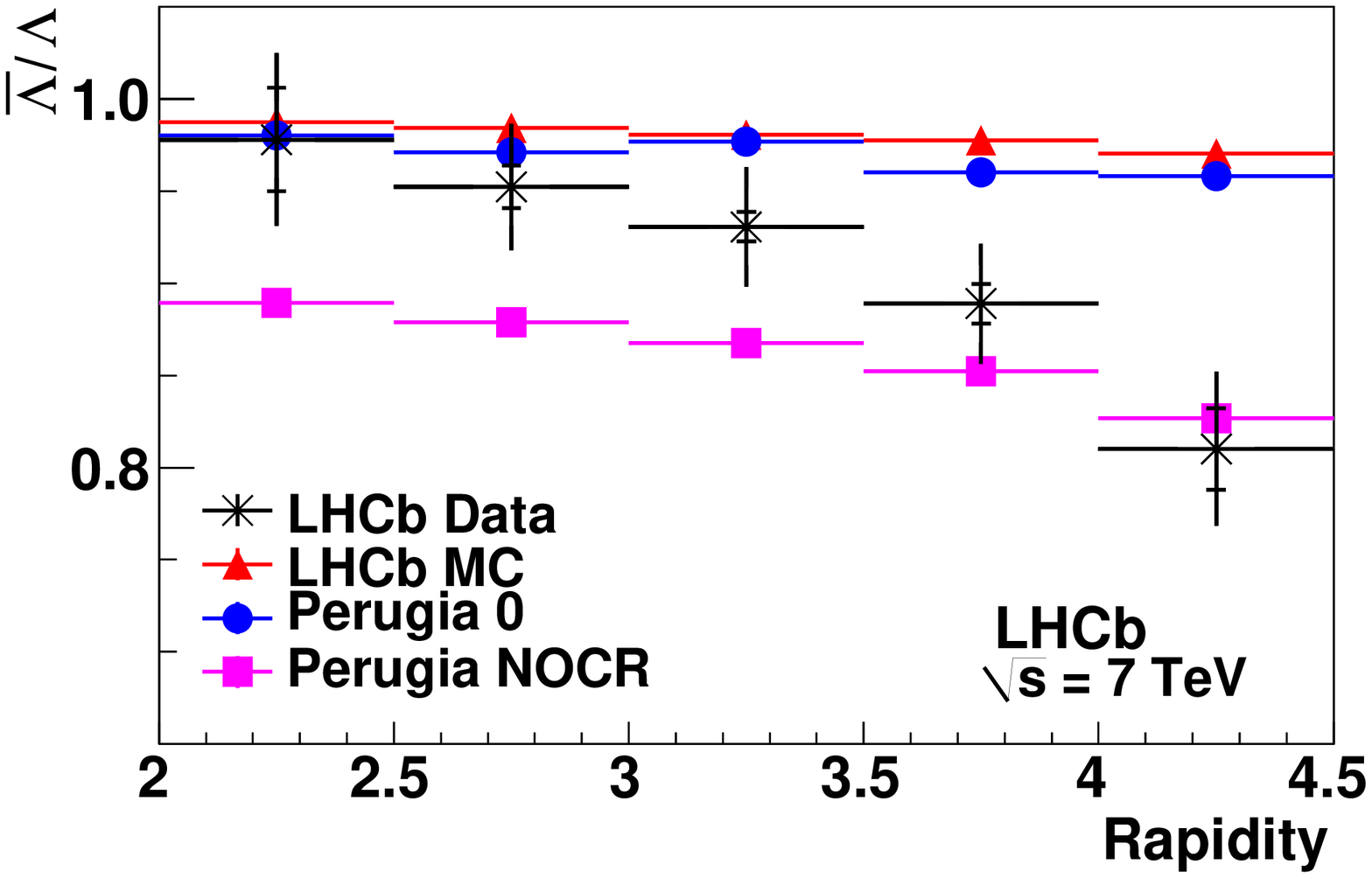}
    \label{fig:plotDataTheory-7-al-y}
  }
  \subfigure[]{
    \includegraphics[width=0.47\textwidth]{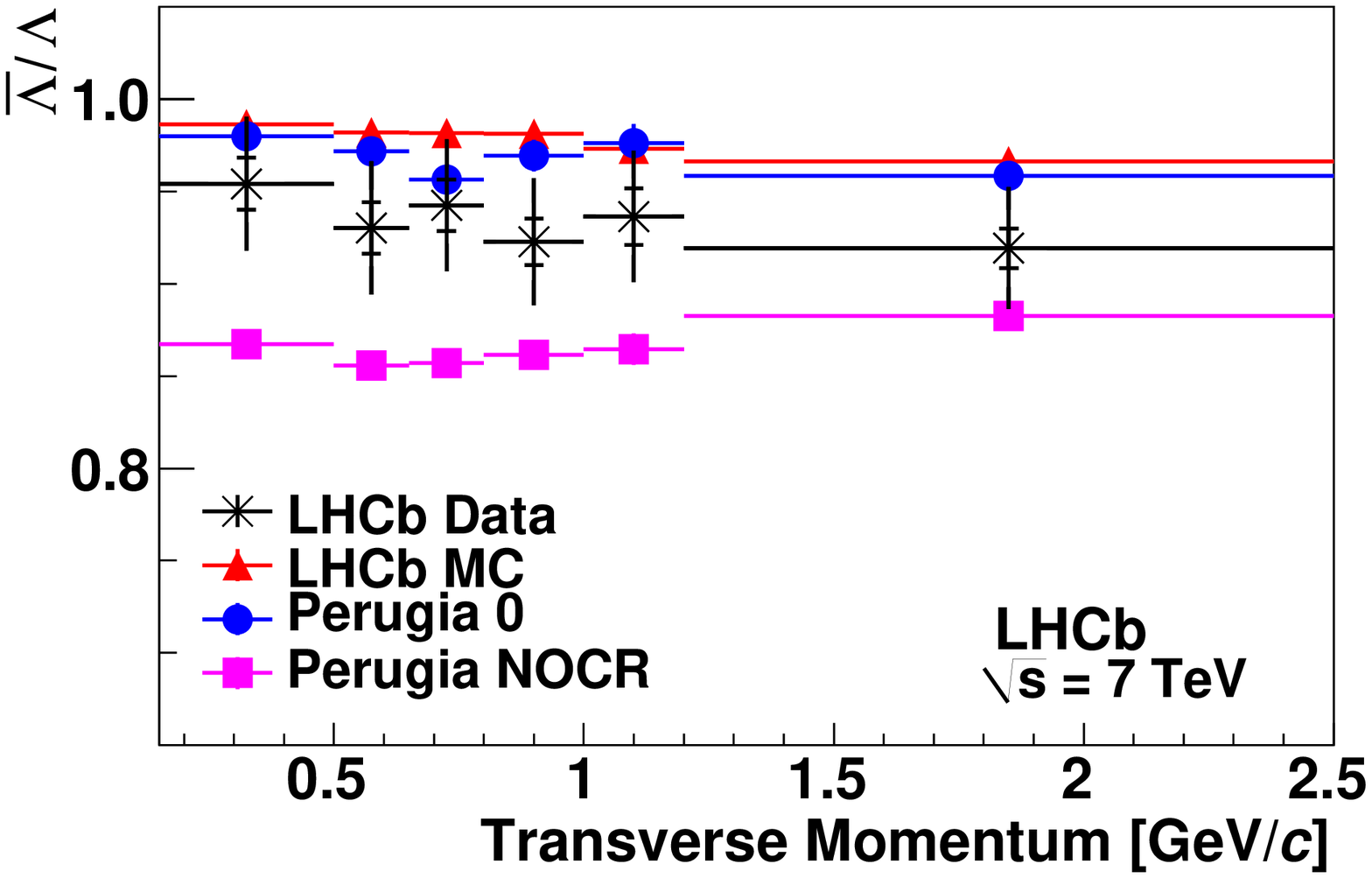}
    \label{fig:plotDataTheory-7-al-pt}
  }
  \subfigure[]{
    \includegraphics[width=0.47\textwidth]{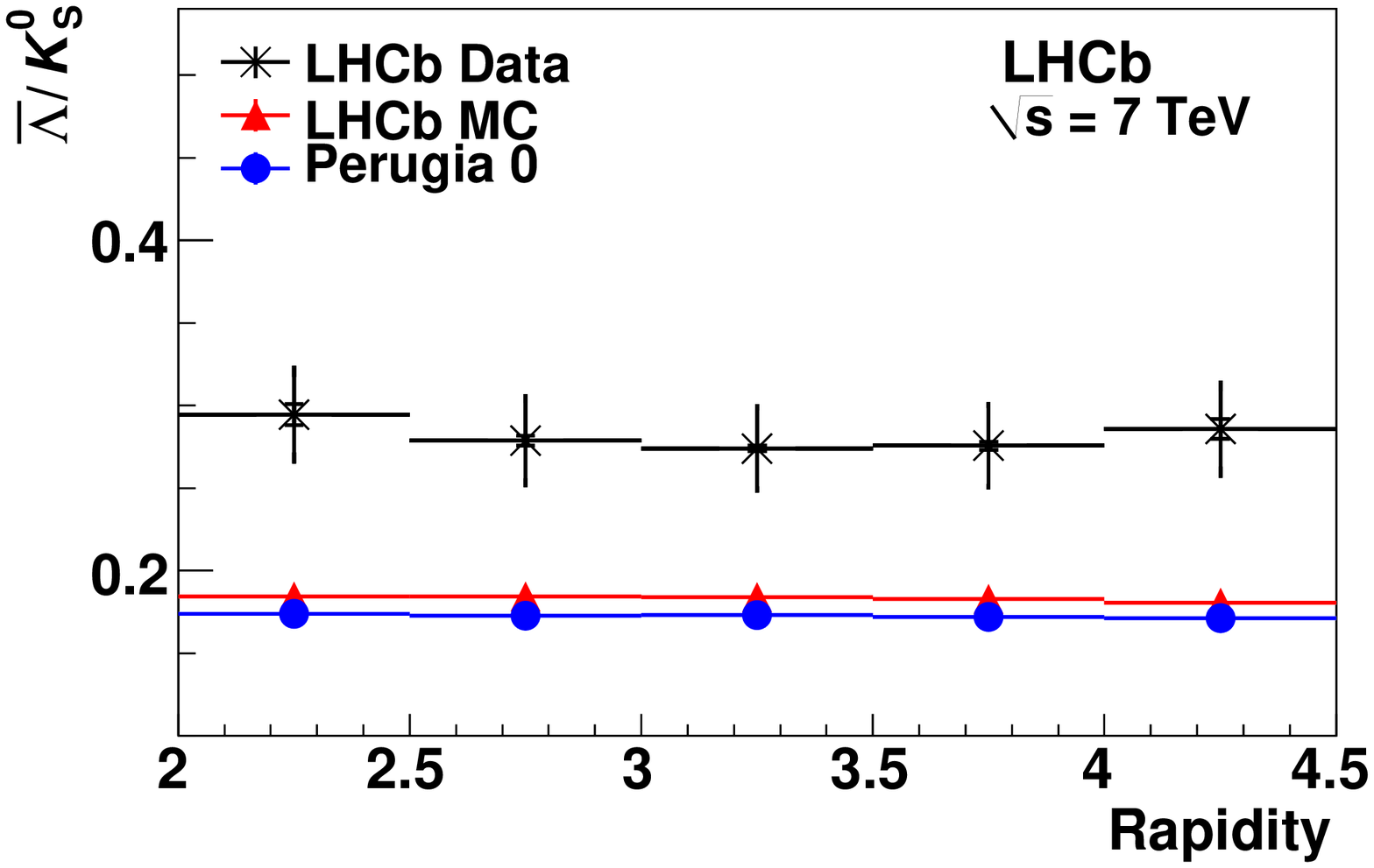}
    \label{fig:plotDataTheory-7-ak-y}
  }
  \subfigure[]{
    \includegraphics[width=0.47\textwidth]{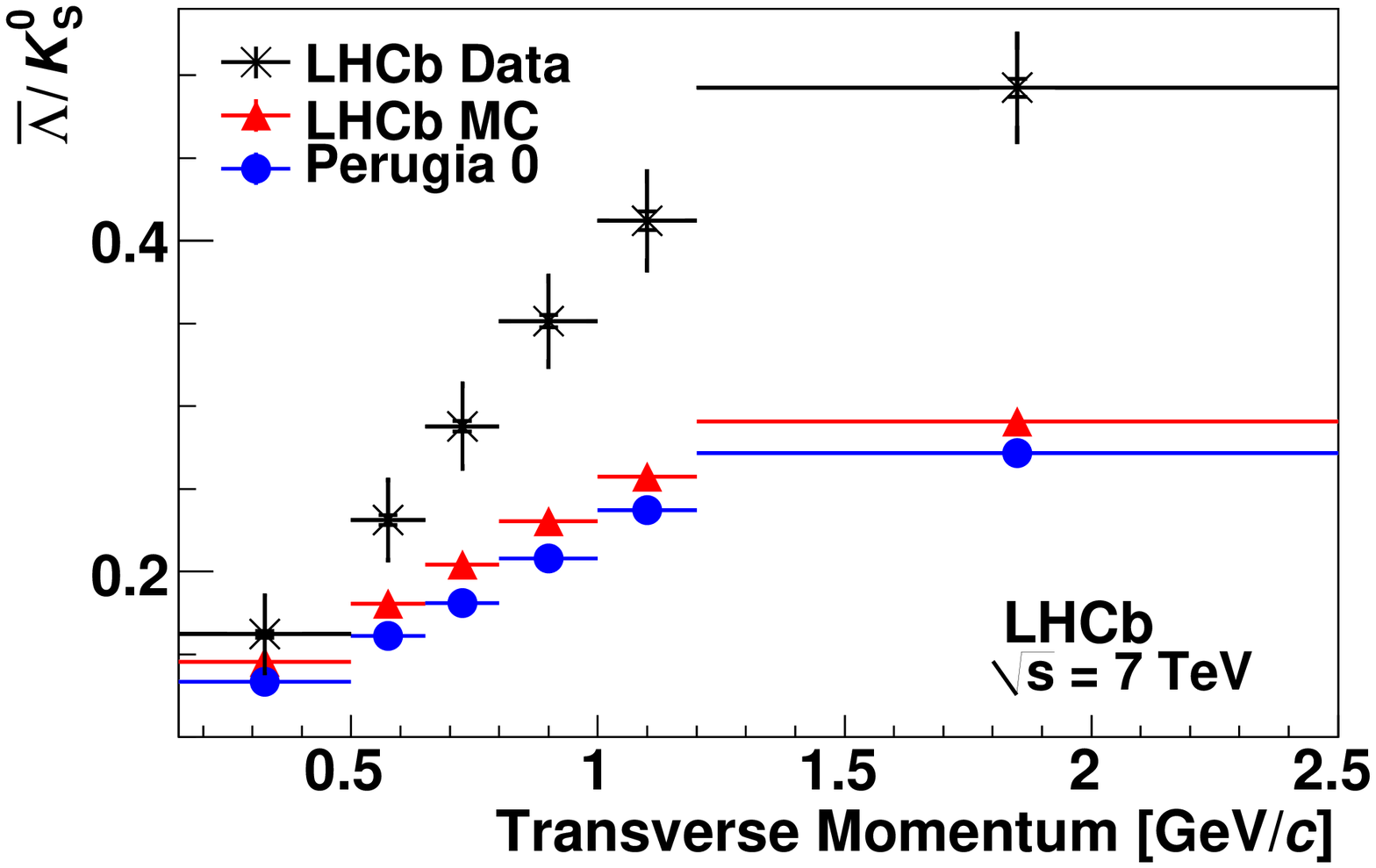}
    \label{fig:plotDataTheory-7-ak-pt}
  }
  \caption{\small The ratios \Lbar/\L and \Lbar/\KS at \LHCHigh compared with the predictions of the LHCb\,MC, Perugia\,0 and Perugia\,NOCR as a function of \subref{fig:plotDataTheory-7-al-y} \& \subref{fig:plotDataTheory-7-ak-y} rapidity and \subref{fig:plotDataTheory-7-al-pt} \& \subref{fig:plotDataTheory-7-ak-pt} transverse momentum.  Vertical lines show the combined statistical and systematic uncertainties and the short horizontal bars (where visible) show the statistical component.}
  \label{fig:plotDataTheory-7}
\end{figure}

\begin{figure}
  \centering
  \subfigure[]{
    \includegraphics[width=0.47\textwidth]{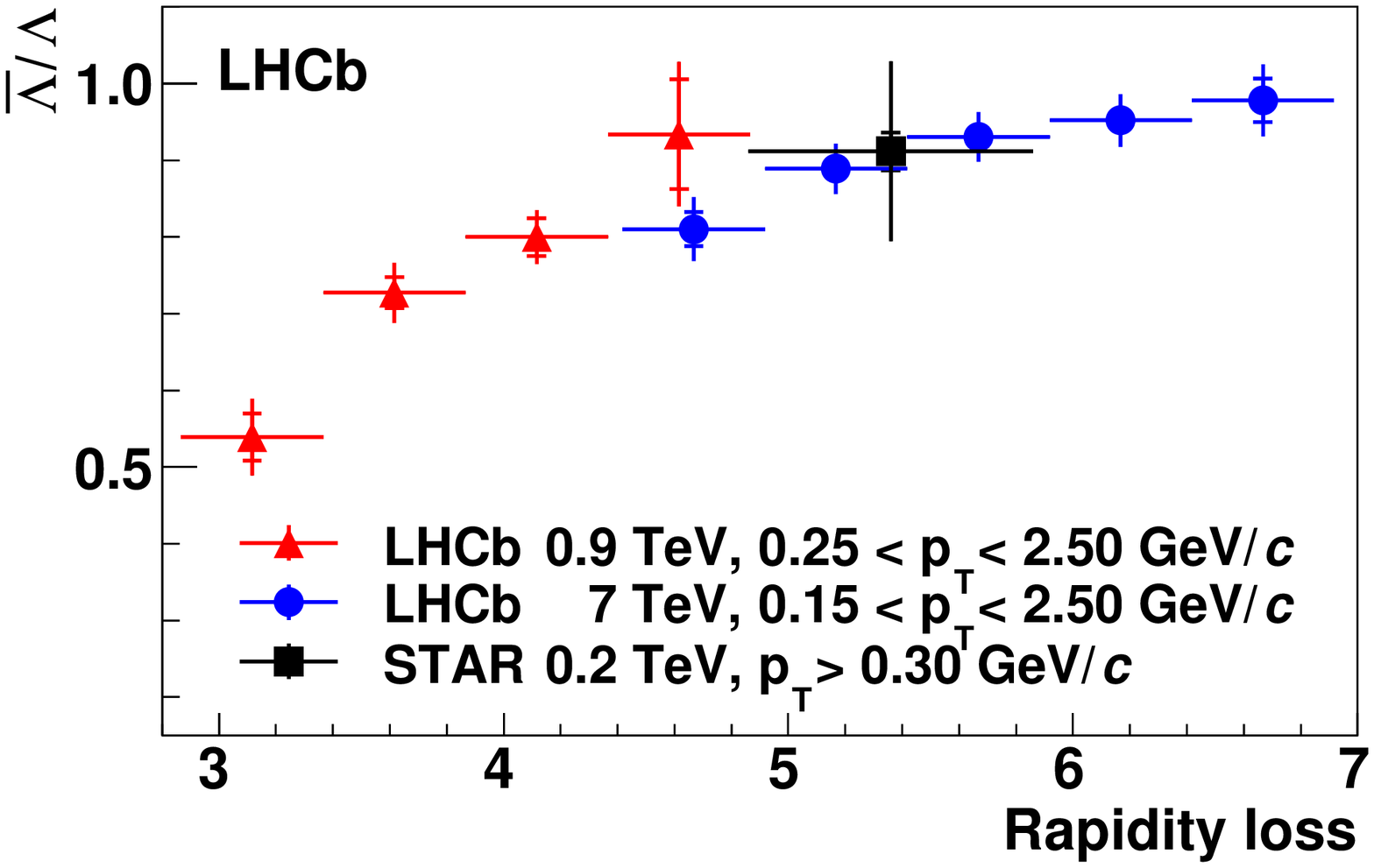}
    \label{fig:plotDeltaY-al}
  }
  \subfigure[]{
    \includegraphics[width=0.47\textwidth]{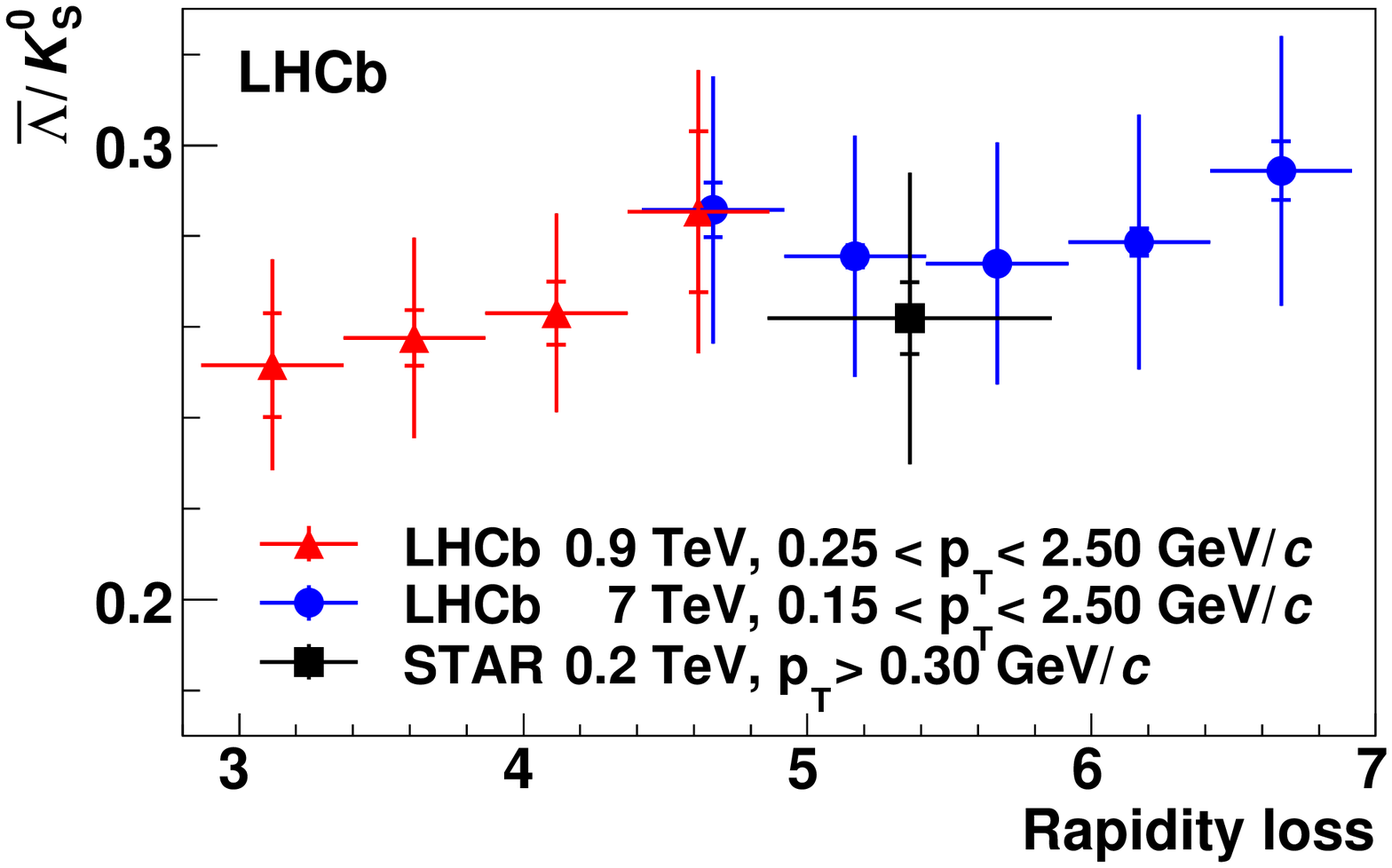}
    \label{fig:plotDeltaY-ak}
  }
  \caption{\small The ratios \subref{fig:plotDeltaY-al} \Lbar/\L and \subref{fig:plotDeltaY-ak} \Lbar/\KS from \lhcb are compared at both \LHCLow (triangles) and 7\,TeV (circles) with the published results from STAR\,\cite{STAR} (squares) as a function of rapidity loss, $\dy = y_\mathrm{beam} - y$.  Vertical lines show the combined statistical and systematic uncertainties and the short horizontal bars (where visible) show the statistical component.}
  \label{fig:plotDeltaY}
\end{figure}

Both measured ratios are compared to the predictions of the \pythia\!6 generator tunes: LHCb MC, Perugia\,0 and Perugia\,NOCR, as functions of \pt and \y at $\s = 0.9\tev$ (Fig.~\ref{fig:plotDataTheory-900}) and at $\s = 7\tev$ (Fig.~\ref{fig:plotDataTheory-7}). According to Monte Carlo studies, as discussed in Section~\ref{sec:systematics}, the requirement for a reconstructed primary vertex results in only a small contribution from diffractive events to the selected \VZ sample, therefore non-diffractive simulated events are used for these comparisons. The predictions of LHCb MC and Perugia\,0 are similar throughout.  The ratio \Lbar/\L is close to Perugia\,0 at low \y but becomes smaller with higher rapidity, approaching Perugia\,NOCR.  In collisions at $\s = 7\tev$, this ratio is consistent with Perugia\,0 across the measured \pt range but is closer to Perugia\,NOCR at $\s = 0.9\tev$.  The production ratio  \Lbar/\KS is larger in data than predicted by Perugia\,0 at both collision energies and in all measurement bins, with the most significant differences observed at high \pt.

To compare results at both collision energies, and to probe scaling violation, both production ratios are shown as a function of rapidity loss, $\dy = y_{\textrm{beam}} - y$, in Fig.~\ref{fig:plotDeltaY}, where $y_{\textrm{beam}}$ is the rapidity of the protons in the anti-clockwise \lhc beam, which travels along the positive $z$ direction through the detector.  Excellent agreement is observed between results at both $\s = 0.9$ and 7\tev as well as with results from \star at $\s = 0.2\tev$.  The measured ratios are also consistent with results published by \alice~\cite{ALICE} and \cms~\cite{CMS}.

The combined field \LHCbUp and \LHCbDown results are also given in tables in Appendix~\ref{app:final-results}.  Results without applying the model dependent non-prompt correction, as discussed in Section~\ref{sec:analysis}, are shown for comparison in Appendix~\ref{app:un-corrected}.

%% -----------------------------------------------------------------
%% Measurement of V0 production ratios in pp collisions at _/s = 0.9 and 7 TeV
%% CERN-PH-EP-2011-082
%% -----------------------------------------------------------------
%% LHCb Collaboration 
%% T. Blake, C. Blanks, W. Bonivento, F. Dettori, R. Muresan
%% c.blanks07@imperial.ac.uk
%% -----------------------------------------------------------------

\section{Conclusions}

The ratio \Lbar/\L is a measurement of the transport of baryon number from \pp collisions to final state hadrons.  There is good agreement with Perugia\,0 at low rapidity which is to be expected since the past experimental results used to test this model have focused on that rapidity region.  At high rapidity however, the measurements favour the extreme baryon transport model of Perugia\,NOCR.  The measured ratio \Lbar/\KS is significantly larger than predicted by Perugia\,0, \ie relatively more baryons are produced in strange hadronisation in data than expected, particularly at higher \pt.  Similar results are found at both $\s = 0.9$ and 7\tev.

When plotted as a function of rapidity loss, \dy, there is excellent agreement between the measurements of both ratios at $\s = 0.9$ and 7\tev as well as with \star's results published at 0.2\tev.  The broad coverage of the measurements in \dy provides a unique data set, which is complementary to previous results.  The \VZ production ratios presented in this paper will help the development of hadronisation models to improve the predictions of Standard Model physics at the \lhc which will define the baseline for new discoveries.

%% -----------------------------------------------------------------
%% Measurement of V0 production ratios in pp collisions at _/s = 0.9 and 7 TeV
%% CERN-PH-EP-2011-082
%% -----------------------------------------------------------------
%% LHCb Collaboration 
%% T. Blake, C. Blanks, W. Bonivento, F. Dettori, R. Muresan
%% c.blanks07@imperial.ac.uk
%% -----------------------------------------------------------------

\section{Acknowledgements}

We express our gratitude to our colleagues in the CERN accelerator departments for the excellent performance of the LHC. We thank the technical and administrative staff at CERN and at the LHCb institutes, and acknowledge support from the National Agencies: CAPES, CNPq, FAPERJ and FINEP (Brazil); CERN; NSFC (China); CNRS/IN2P3 (France); BMBF, DFG, HGF and MPG (Germany); SFI (Ireland); INFN (Italy); FOM and NWO (Netherlands); SCSR (Poland); ANCS (Romania); MinES of Russia and Rosatom (Russia); MICINN, XUNGAL and GENCAT (Spain); SNSF and SER (Switzerland); NAS Ukraine (Ukraine); STFC (United Kingdom); NSF (USA).  We also acknowledge the support received from the ERC under FP7 and the R$\mathrm{\acute{e}}$gion Auvergne.

\clearpage
%\bibliographystyle{LHCb}
%\bibliography{references}
%% -----------------------------------------------------------------
%% Measurement of V0 production ratios in pp collisions at _/s = 0.9 and 7 TeV
%% CERN-PH-EP-2011-082
%% -----------------------------------------------------------------
%% LHCb Collaboration 
%% T. Blake, C. Blanks, W. Bonivento, F. Dettori, R. Muresan
%% c.blanks07@imperial.ac.uk
%% -----------------------------------------------------------------

\providecommand{\href}[2]{#2}\begingroup\raggedright\endgroup

%% -----------------------------------------------------------------
%% Measurement of V0 production ratios in pp collisions at _/s = 0.9 and 7 TeV
%% CERN-PH-EP-2011-082
%% -----------------------------------------------------------------
%% LHCb Collaboration 
%% T. Blake, C. Blanks, W. Bonivento, F. Dettori, R. Muresan
%% c.blanks07@imperial.ac.uk
%% -----------------------------------------------------------------

\newpage
{\noindent\bf\Large Appendix}

\appendix

\section{Tabulated results}
\label{app:final-results}

\begin{table}[h!]
  \centering
 \caption{\small The production ratios \Lbar/\L and \Lbar/\KS, measured at \LHCLow, are quoted in percent with statistical and systematic errors as a function of \subref{tab:val-stat-syst-900-al} \& \subref{tab:val-stat-syst-900-ak} rapidity, \y, and \subref{tab:val-stat-syst-900-pt} transverse momentum, \pt\!$[\!\gevc]$.}

  \small
  \subtable[]{
    \begin{tabular}{ccccc}
\toprule
$ \Lbar/\L $ & $2.0<y<2.5$ & $2.5<y<3.0$ & $3.0<y<3.5$ & $3.5<y<4.0$ \\
\midrule
$0.25<\pt<2.50$ & 93.4$\pm$7.2$\pm$6.1 & 80.0$\pm$2.5$\pm$2.5 & 72.7$\pm$2.0$\pm$3.3 & 53.9$\pm$3.1$\pm$4.0 \\
\midrule
$0.25<\pt<0.65$ & 162.2$\pm$48.2$\pm$6.6 & 90.4$\pm$6.6$\pm$3.0 & 61.0$\pm$4.2$\pm$3.5 & 42.0$\pm$12.4$\pm$5.3 \\
$0.65<\pt<1.00$ & 72.3$\pm$9.7$\pm$2.5 & 77.2$\pm$3.9$\pm$2.4 & 74.6$\pm$3.3$\pm$3.9 & 61.7$\pm$5.6$\pm$3.6 \\
$1.00<\pt<2.50$ & 90.4$\pm$11.3$\pm$2.8 & 74.5$\pm$4.6$\pm$2.4 & 75.7$\pm$3.4$\pm$3.1 & 48.5$\pm$3.8$\pm$2.2 \\
\bottomrule
    \end{tabular}
    \label{tab:val-stat-syst-900-al}
  }
  \subtable[]{
    \begin{tabular}{ccccc}
\toprule
$ \Lbar/\KS $ & $2.0<y<2.5$ & $2.5<y<3.0$ & $3.0<y<3.5$ & $3.5<y<4.0$ \\ 
\midrule
$0.25<\pt<2.50$ & 28.5$\pm$1.8$\pm$2.6 & 26.3$\pm$0.7$\pm$2.1 & 25.8$\pm$0.6$\pm$2.1 & 25.2$\pm$1.1$\pm$2.0 \\
\midrule
$0.25<\pt<0.65$ & 19.7$\pm$3.6$\pm$2.6 & 21.8$\pm$1.4$\pm$2.2 & 18.0$\pm$1.0$\pm$1.8 & 15.8$\pm$3.1$\pm$2.1 \\
$0.65<\pt<1.00$ & 31.6$\pm$2.9$\pm$2.5 & 30.6$\pm$1.3$\pm$2.3 & 30.0$\pm$1.2$\pm$2.2 & 29.9$\pm$2.1$\pm$2.2 \\
$1.00<\pt<2.50$ & 46.3$\pm$4.5$\pm$2.9 & 42.9$\pm$2.1$\pm$2.5 & 41.3$\pm$1.6$\pm$3.2 & 32.3$\pm$2.0$\pm$2.6 \\
\bottomrule
    \end{tabular}
    \label{tab:val-stat-syst-900-ak}
  }
  \subtable[]{
    \begin{tabular}{ccc}
\toprule
   $ 2.0<y<4.0 $ & \Lbar/\L & \Lbar/\KS \\ 
\midrule
$0.25<\pt<0.50$ & 80.6$\pm$4.6$\pm$4.0 & 17.7$\pm$0.8$\pm$1.7 \\
$0.50<\pt<0.65$ & 73.1$\pm$3.6$\pm$3.2 & 21.8$\pm$0.9$\pm$1.8 \\
$0.65<\pt<0.80$ & 73.7$\pm$3.2$\pm$3.7 & 28.4$\pm$1.0$\pm$2.3 \\
$0.80<\pt<1.00$ & 77.5$\pm$3.2$\pm$3.7 & 32.3$\pm$1.2$\pm$2.4 \\
$1.00<\pt<1.20$ & 70.1$\pm$3.4$\pm$2.3 & 36.8$\pm$1.5$\pm$2.4 \\
$1.20<\pt<2.50$ & 74.5$\pm$3.0$\pm$2.5 & 44.2$\pm$1.5$\pm$2.8 \\
\bottomrule
    \end{tabular}
    \label{tab:val-stat-syst-900-pt}
  }
\label{tab:val-stat-syst-900}
\end{table}

\clearpage

\begin{table}[h!]
  \centering
 \caption{\small The production ratios \Lbar/\L and \Lbar/\KS, measured at \LHCHigh, are quoted in percent with statistical and systematic errors as a function of \subref{tab:val-stat-syst-7-al} \& \subref{tab:val-stat-syst-7-ak} rapidity, \y, and \subref{tab:val-stat-syst-7-pt} transverse momentum, \pt\!$[\!\gevc]$.}

  \small
  \subtable[]{
    \begin{tabular}{cccccc}
\toprule
$ \Lbar/\L $ & $2.0<y<2.5$ & $2.5<y<3.0$ & $3.0<y<3.5$ & $3.5<y<4.0$ & $4.0<y<4.5$ \\ 
\midrule
$0.15<\pt<2.50$ & 97.8$\pm$2.8$\pm$3.8 & 95.2$\pm$1.2$\pm$3.2 & 93.1$\pm$0.8$\pm$3.1 & 88.9$\pm$1.1$\pm$3.1 & 81.0$\pm$2.2$\pm$3.5 \\
\midrule
$0.15<\pt<0.65$ & 87.2$\pm$16.7$\pm$11.0 & 95.7$\pm$1.8$\pm$3.5 & 94.2$\pm$1.4$\pm$3.3 & 87.6$\pm$2.3$\pm$3.2 & 90.0$\pm$12.6$\pm$4.2 \\
$0.65<\pt<1.00$ & 97.4$\pm$5.3$\pm$3.9 & 96.8$\pm$2.2$\pm$3.5 & 92.4$\pm$1.3$\pm$3.3 & 89.6$\pm$1.8$\pm$3.2 & 86.2$\pm$4.2$\pm$3.2 \\
$1.00<\pt<2.50$ & 98.7$\pm$2.9$\pm$3.4 & 96.6$\pm$1.8$\pm$3.3 & 92.8$\pm$1.5$\pm$3.2 & 90.3$\pm$1.7$\pm$3.2 & 79.2$\pm$2.8$\pm$2.9 \\
\bottomrule
    \end{tabular}
    \label{tab:val-stat-syst-7-al}
  }
  \subtable[]{
    \begin{tabular}{cccccc}
\toprule
$ \Lbar/\KS $ & $2.0<y<2.5$ & $2.5<y<3.0$ & $3.0<y<3.5$ & $3.5<y<4.0$ & $4.0<y<4.5$ \\ 
\midrule
$0.15<\pt<2.50$ & 29.4$\pm$0.6$\pm$2.9 & 27.9$\pm$0.3$\pm$2.8 & 27.4$\pm$0.2$\pm$2.7 & 27.6$\pm$0.3$\pm$2.6 & 28.6$\pm$0.6$\pm$2.9 \\
\midrule
$0.15<\pt<0.65$ & 18.2$\pm$2.7$\pm$3.0 & 19.1$\pm$0.3$\pm$2.6 & 18.5$\pm$0.2$\pm$2.5 & 17.5$\pm$0.4$\pm$2.5 & 20.7$\pm$1.5$\pm$3.0 \\
$0.65<\pt<1.00$ & 32.0$\pm$1.3$\pm$3.0 & 32.8$\pm$0.6$\pm$3.0 & 31.5$\pm$0.4$\pm$2.8 & 29.9$\pm$0.5$\pm$2.8 & 32.1$\pm$1.2$\pm$2.9 \\
$1.00<\pt<2.50$ & 48.3$\pm$1.1$\pm$3.5 & 47.8$\pm$0.7$\pm$3.3 & 45.8$\pm$0.6$\pm$3.3 & 45.6$\pm$0.7$\pm$3.2 & 39.9$\pm$1.0$\pm$3.0 \\
\bottomrule
    \end{tabular}
    \label{tab:val-stat-syst-7-ak}
  }
  \subtable[]{
    \begin{tabular}{ccc}
\toprule
   $2.0<y<4.5$ & \Lbar/\L & \Lbar/\KS \\ 
\midrule
$0.15<\pt<0.50$ & 95.4$\pm$1.4$\pm$3.4 & 16.2$\pm$0.2$\pm$2.4 \\
$0.50<\pt<0.65$ & 93.0$\pm$1.4$\pm$3.3 & 23.1$\pm$0.3$\pm$2.5 \\
$0.65<\pt<0.80$ & 94.3$\pm$1.4$\pm$3.3 & 28.8$\pm$0.3$\pm$2.7 \\
$0.80<\pt<1.00$ & 92.3$\pm$1.3$\pm$3.2 & 35.1$\pm$0.4$\pm$2.8 \\
$1.00<\pt<1.20$ & 93.6$\pm$1.5$\pm$3.2 & 41.2$\pm$0.6$\pm$3.0 \\
$1.20<\pt<2.50$ & 91.9$\pm$1.1$\pm$3.1 & 49.2$\pm$0.5$\pm$3.4 \\
\bottomrule
    \end{tabular}
    \label{tab:val-stat-syst-7-pt}
  }
 \label{tab:val-stat-syst-7}
\end{table}

%%%%%%%%%%%%%%%%%%%%%%%%%%%%%%%%%%%%%%%%%%%%%%%%%%%%%%%%5

\clearpage

\section{Tabulated results before non-prompt correction}
\label{app:un-corrected}

\begin{table}[h!]
  \centering
  \caption{\small The production ratios \Lbar/\L and \Lbar/\KS without non-prompt corrections at \LHCLow are quoted in percent with statistical and systematic errors as a function of \subref{tab:no-correction-900-al} \& \subref{tab:no-correction-900-ak}  rapidity, \y, and \subref{tab:no-correction-900-pt} transverse momentum, \pt\!$[\!\gevc]$.}

  \small
  \subtable[]{
    \begin{tabular}{ccccc}
\toprule
$ \Lbar/\L $ & $2.0<y<2.5$ & $2.5<y<3.0$ & $3.0<y<3.5$ & $3.5<y<4.0$ \\
\midrule
$0.25<\pt<2.50$ & 93.1$\pm$7.2$\pm$6.0 & 79.3$\pm$2.5$\pm$2.4 & 73.2$\pm$2.0$\pm$3.2 & 54.1$\pm$3.1$\pm$3.9 \\
\midrule
$0.25<\pt<0.65$ & 163.7$\pm$48.2$\pm$6.5 & 89.2$\pm$6.6$\pm$2.8 & 61.5$\pm$4.2$\pm$3.4 & 41.4$\pm$12.4$\pm$5.3 \\
$0.65<\pt<1.00$ & 71.8$\pm$9.7$\pm$2.4 & 76.5$\pm$3.9$\pm$2.2 & 75.2$\pm$3.3$\pm$3.8 & 62.0$\pm$5.6$\pm$3.5 \\
$1.00<\pt<2.50$ & 89.9$\pm$11.3$\pm$2.7 & 74.2$\pm$4.6$\pm$2.3 & 75.7$\pm$3.4$\pm$3.0 & 48.5$\pm$3.8$\pm$2.1 \\
\bottomrule
    \end{tabular}
    \label{tab:no-correction-900-al}
  }
  \subtable[]{
    \begin{tabular}{ccccc}
\toprule
$ \Lbar/\KS $ & $2.0<y<2.5$ & $2.5<y<3.0$ & $3.0<y<3.5$ & $3.5<y<4.0$ \\ 
\midrule
$0.25<\pt<2.50$ & 28.9$\pm$1.8$\pm$2.4 & 27.2$\pm$0.7$\pm$1.9 & 26.6$\pm$0.6$\pm$1.9 & 25.6$\pm$1.1$\pm$1.8 \\
\midrule
$0.25<\pt<0.65$ & 20.7$\pm$3.6$\pm$2.4 & 23.0$\pm$1.4$\pm$2.0 & 18.9$\pm$1.0$\pm$1.6 & 16.3$\pm$3.1$\pm$1.9 \\
$0.65<\pt<1.00$ & 31.9$\pm$2.9$\pm$2.3 & 31.5$\pm$1.3$\pm$2.1 & 31.0$\pm$1.2$\pm$2.0 & 30.6$\pm$2.1$\pm$2.0 \\
$1.00<\pt<2.50$ & 46.7$\pm$4.5$\pm$2.8 & 43.1$\pm$2.1$\pm$2.4 & 41.9$\pm$1.6$\pm$3.0 & 32.5$\pm$2.0$\pm$2.4 \\
\bottomrule
    \end{tabular}
    \label{tab:no-correction-900-ak}
  }
  \subtable[]{
    \begin{tabular}{ccc}
\toprule
   $ 2.0<y<4.0 $ & \Lbar/\L & \Lbar/\KS \\ 
\midrule
$0.25<\pt<0.50$ & 80.1$\pm$4.6$\pm$3.9 & 18.8$\pm$0.8$\pm$1.5 \\
$0.50<\pt<0.65$ & 72.9$\pm$3.6$\pm$3.1 & 22.9$\pm$0.9$\pm$1.6 \\
$0.65<\pt<0.80$ & 73.9$\pm$3.2$\pm$3.6 & 29.5$\pm$1.0$\pm$2.1 \\
$0.80<\pt<1.00$ & 77.5$\pm$3.2$\pm$3.5 & 33.1$\pm$1.2$\pm$2.3 \\
$1.00<\pt<1.20$ & 70.1$\pm$3.4$\pm$2.1 & 37.2$\pm$1.5$\pm$2.2 \\
$1.20<\pt<2.50$ & 74.4$\pm$3.0$\pm$2.3 & 44.5$\pm$1.5$\pm$2.6 \\
\bottomrule
    \end{tabular}
    \label{tab:no-correction-900-pt}
  }
\label{tab:no-correction-900}
\end{table}

\newpage

\begin{table}[h!]
  \centering
  \caption{\small The production ratios \Lbar/\L and \Lbar/\KS without non-prompt corrections at \LHCHigh are quoted in percent with statistical and systematic errors as a function of \subref{tab:no-correction-7-al} \& \subref{tab:no-correction-7-ak} rapidity, \y, and \subref{tab:no-correction-7-pt} transverse momentum, \pt\!$[\!\gevc]$.}

  \small
  \subtable[]{
    \begin{tabular}{cccccc}
\toprule
$ \Lbar/\L $ & $2.0<y<2.5$ & $2.5<y<3.0$ & $3.0<y<3.5$ & $3.5<y<4.0$ & $4.0<y<4.5$ \\ 
\midrule
$0.15<\pt<2.50$ & 97.3$\pm$2.8$\pm$3.6 & 95.1$\pm$1.2$\pm$3.1 & 92.7$\pm$0.8$\pm$3.0 & 88.6$\pm$1.1$\pm$2.9 & 80.9$\pm$2.2$\pm$3.4 \\
\midrule
$0.15<\pt<0.65$ & 85.6$\pm$16.7$\pm$11.0 & 95.4$\pm$1.8$\pm$3.4 & 93.9$\pm$1.4$\pm$3.2 & 87.3$\pm$2.3$\pm$3.1 & 90.1$\pm$12.6$\pm$4.1 \\
$0.65<\pt<1.00$ & 97.5$\pm$5.3$\pm$3.8 & 96.5$\pm$2.2$\pm$3.4 & 91.8$\pm$1.3$\pm$3.1 & 89.5$\pm$1.8$\pm$3.1 & 86.2$\pm$4.2$\pm$3.0 \\
$1.00<\pt<2.50$ & 98.2$\pm$2.9$\pm$3.3 & 96.6$\pm$1.8$\pm$3.2 & 92.5$\pm$1.5$\pm$3.1 & 90.0$\pm$1.7$\pm$3.1 & 79.0$\pm$2.8$\pm$2.8 \\
\bottomrule
    \end{tabular}
    \label{tab:no-correction-7-al}
  }
  \subtable[]{
    \begin{tabular}{cccccc}
\toprule
$ \Lbar/\KS $ & $2.0<y<2.5$ & $2.5<y<3.0$ & $3.0<y<3.5$ & $3.5<y<4.0$ & $4.0<y<4.5$ \\ 
\midrule
$0.15<\pt<2.50$ & 29.4$\pm$0.6$\pm$2.8 & 28.4$\pm$0.3$\pm$2.6 & 28.0$\pm$0.2$\pm$2.5 & 27.9$\pm$0.3$\pm$2.5 & 28.7$\pm$0.6$\pm$2.7 \\
\midrule
$0.15<\pt<0.65$ & 18.5$\pm$2.7$\pm$2.9 & 20.0$\pm$0.3$\pm$2.5 & 19.2$\pm$0.2$\pm$2.3 & 17.9$\pm$0.4$\pm$2.3 & 21.1$\pm$1.5$\pm$2.9 \\
$0.65<\pt<1.00$ & 32.3$\pm$1.3$\pm$2.9 & 33.3$\pm$0.6$\pm$2.8 & 32.2$\pm$0.4$\pm$2.7 & 30.2$\pm$0.5$\pm$2.6 & 32.2$\pm$1.2$\pm$2.7 \\
$1.00<\pt<2.50$ & 47.9$\pm$1.1$\pm$3.3 & 47.5$\pm$0.7$\pm$3.2 & 45.7$\pm$0.6$\pm$3.2 & 45.6$\pm$0.7$\pm$3.1 & 39.5$\pm$1.0$\pm$2.8 \\
\bottomrule
    \end{tabular}
    \label{tab:no-correction-7-ak}
  }
  \subtable[]{
    \begin{tabular}{ccc}
\toprule
   $ 2.0<y<4.5 $ & \Lbar/\L & \Lbar/\KS \\ 
\midrule
$0.15<\pt<0.50$ & 95.0$\pm$1.4$\pm$3.2 & 16.9$\pm$0.2$\pm$2.3 \\
$0.50<\pt<0.65$ & 92.9$\pm$1.4$\pm$3.2 & 23.8$\pm$0.3$\pm$2.4 \\
$0.65<\pt<0.80$ & 94.0$\pm$1.4$\pm$3.2 & 29.4$\pm$0.3$\pm$2.5 \\
$0.80<\pt<1.00$ & 91.9$\pm$1.3$\pm$3.1 & 35.5$\pm$0.4$\pm$2.7 \\
$1.00<\pt<1.20$ & 93.1$\pm$1.5$\pm$3.1 & 41.3$\pm$0.6$\pm$2.9 \\
$1.20<\pt<2.50$ & 91.8$\pm$1.1$\pm$3.0 & 48.9$\pm$0.5$\pm$3.2 \\
\bottomrule
    \end{tabular}
    \label{tab:no-correction-7-pt}
  }
 \label{tab:no-correction-7}
\end{table}

\end{document}